\documentclass[%
 superscriptaddress,
 amsmath,amssymb,
 aps,
 prab,
twocolumn,
 floatfix,
]{revtex4-2}

\usepackage{graphicx}
\usepackage{dcolumn}
\usepackage{bm}
\usepackage{hyperref}
\hypersetup{
    colorlinks=true,
    citecolor=blue,
    linkcolor=blue,
    urlcolor=blue
    }

\usepackage[euler]{textgreek}
\usepackage{makecell}
\usepackage{multirow}

\let\oldhash\#
\renewcommand{\#}{\texttt{\oldhash}} 

\usepackage{lineno}

\usepackage{graphicx}
\usepackage{dcolumn}
\usepackage{bm}
\usepackage{adjustbox}
\usepackage{subfigure}
\usepackage{epstopdf}
\usepackage{amsmath}
\usepackage{lineno,hyperref}
\usepackage{makecell}
\usepackage{multirow}
\usepackage{rotating}
\usepackage{enumitem}
\usepackage{url}
\usepackage{color,soul}
\modulolinenumbers[5]

\begin{document}


\title{Design development and implementation of an irradiation station at the neutron time-of-flight facility at CERN}

\author{M.~Ferrari}
\email{matteo.ferrari.2@cern.ch}
\affiliation{European Organization for Nuclear Research, CERN, Esplanade des Particules 1, 1211 Geneve 23, Switzerland}
\author{D.~Senajova}
\affiliation{European Organization for Nuclear Research, CERN, Esplanade des Particules 1, 1211 Geneve 23, Switzerland}
\affiliation{Imperial College London, SW7 2AZ, London, United Kingdom}
\author{O.~Aberle}
\author{Y.~Q.~Aguiar}
\author{D.~Baillard}
\author{M.~Barbagallo}
\altaffiliation{Currently at Transmutex, Geneva, Switzerland.}
\affiliation{European Organization for Nuclear Research, CERN, Esplanade des Particules 1, 1211 Geneve 23, Switzerland}
\author{A.-P.~Bernardes}
\author{L.~Buonocore}
\author{M.~Cecchetto}
\author{V.~Clerc}
\author{M.~Di~Castro}
\author{R.~Garcia~Alia}
\author{S.~Girod}
\author{J.-L.~Grenard}
\author{K.~Kershaw}
\author{G.~Lerner}
\author{M.~Maeder}
\altaffiliation{Currently at the University of Applied Sciences Offenburg, Department of Electrical Engineering, Medical Engineering and Computer Science, Badstraße 24, 77652 Offenburg, Germany.}
\affiliation{European Organization for Nuclear Research, CERN, Esplanade des Particules 1, 1211 Geneve 23, Switzerland}
\author{A.~Makovec}
\affiliation{European Organization for Nuclear Research, CERN, Esplanade des Particules 1, 1211 Geneve 23, Switzerland}
\author{A.~Mengoni}
\affiliation{European Organization for Nuclear Research, CERN, Esplanade des Particules 1, 1211 Geneve 23, Switzerland}
\affiliation{ENEA, Italian National Agency for New Technologies, Energy and Sustainable Economic Development, Italy}
\author{M.~Perez~Ornedo}
\author{F.~Pozzi}
\affiliation{European Organization for Nuclear Research, CERN, Esplanade des Particules 1, 1211 Geneve 23, Switzerland}
\author{C.~V.~Almagro}
\affiliation{European Organization for Nuclear Research, CERN, Esplanade des Particules 1, 1211 Geneve 23, Switzerland}
\affiliation{Jaume I University of Castellon, 12006 Castellon de la Plana, Spain}
\author{M.~Calviani}
\email{marco.calviani@cern.ch}
\affiliation{European Organization for Nuclear Research, CERN, Esplanade des Particules 1, 1211 Geneve 23, Switzerland}

\collaboration{for the n\_TOF Collaboration}

\date{\today}

\begin{abstract}

A new parasitic, mixed-field, neutron-dominated irradiation station has been recently commissioned at the European Laboratory for Particle Physics (CERN). The station is installed within the Neutron Time-Of-Flight (n\_TOF) facility, taking advantage of the secondary radiation produced by the neutron spallation target, with neutrons ranging from 0.025~eV to several hundreds of MeV. 

The new station allows radiation damage studies to be performed in irradiation conditions that are closer to the ones encountered during the operation of particle accelerators; the irradiation tests carried out in the station will be complementary to the standard tests on materials, usually performed with gamma sources. 

Samples will be exposed to neutron-dominated doses in the MGy range per year, with minimal impact on the n\_TOF facility operation. The station has twenty-four irradiation positions, each hosting up to 100~cm$^3$ of sample material.  

In view of its proximity to the n\_TOF target, inside protective shielding, the irradiation station and its operating procedures have  been carefully developed taking into account the safety of personnel and to avoid any unwanted impact on the operation of the n\_TOF facility and experiments. Due to the residual radioactivity of the whole area around the n\_TOF target and of the irradiated samples, access to the irradiation station is forbidden to human operators even when the n\_TOF facility is not in operation. Robots are used for the remote installation and retrieval of the samples, and other optimizations of the handling procedures were developed in compliance with radiation protection regulations and the aim of minimizing doses to personnel. 

The sample containers were designed to be radiation tolerant, compatible with remote handling and subject to detailed risk analysis and testing during their development. The whole life cycle of the irradiated materials, including their post-irradiation examinations and final disposal was considered and optimized. 

\end{abstract}

\pacs{}

\maketitle







\section{Introduction}
\label{sec:intro}

Despite their known sensitivity to radiation \cite{bolt,seguchi}, non-metallic materials are, of necessity, extensively used in high-radiation areas of accelerators and high-power physics facilities~\cite{ITER,ferrari2021book}. Commercial polymeric components such as lubricating oils, greases, elastomeric O-rings, insulators and optical fibers are used in crucial equipment in areas where intense levels of radiation are encountered at research laboratories worldwide. Examples of large infrastructures where the selection of radiation tolerant materials is critical for successful operation include: the high intensity proton accelerator facility at the Japan Proton Accelerator Research Complex (J-PARC, Japan), the accelerator complex at Fermilab (USA), the Spallation Neutron Source at the Oak Ridge National Laboratory (ORNL, USA), the ITER collaboration (France), the European Spallation Source (ESS, Sweden) as well as the accelerator complex at the European Laboratory of Particle Physics (CERN, Switzerland). Additionally, various other types of more compact high-power target facilities currently operating or under development, such as ISOLDE (Isotope Separation On-Line) facilities~\cite{Catherall_2017,isol} and MEDICIS (Medical Isotopes Collected from ISOLDE)~\cite{10.3389/fmed.2021.693682}, share comparable challenges in the selection of materials to be used for their construction and operation.

Radiation-induced degradation of specific components can limit the lifetime of accelerator equipment such as Beam Intercepting Devices  (BIDs)~\cite{ferrari2021,esposito2021,maestre2021}, including complex high-power target assemblies, necessitating their earlier replacement~\cite{ISOLDE,giles2020}. In the above-mentioned devices, commercial materials are typically exposed to mixed radiation fields, absorbing physical doses ranging up to several~MGy during their lifetime~\cite{maestre2021,ferrari2021}. Many types of polymeric materials are reported to fail at comparable doses~\cite{bolt}. 

Historically, accelerated irradiation tests of commercial materials have been performed for the development of nuclear, aerospace, fusion and accelerator technologies~\cite{ITER,ITERreport,georgia,rice,nasa,IAEC}. At CERN, extensive tests of materials and components to be used in accelerators have been performed between the 1960's and the early 2000's. The results are reported in a series of documents known as Yellow Reports; examples are given in References~\cite{yellow1, yellow2,yellow3}. Most of the results have been collected using gamma sources, under the assumption that equal doses roughly induce equal damage, regardless of the radiation type~\cite{yellow3,seguchi}.

However, recent studies~\cite{briskman,rivaton,ferrari2021oil} question this hypothesis and report a general lack of up-to-date scientific knowledge in this field~\cite{IAEAmeeting}, highlighting the need for new data to be collected under different irradiation conditions. Due to the continuous upgrading and development of accelerators and to the increased intensity of the produced radiation, new irradiation studies are necessary to correspond with the more demanding requirements on materials for use in accelerators and to minimize radiation-induced failures~\cite{garcia2018,hiradmat}.

For example, irradiation tests on materials using gamma sources are generally accelerated so that total doses comparable to the ones absorbed in years or decades of operation can be delivered in much shorter times in testing conditions. Data collected in mixed fields and longer irradiation times are currently very scarce, but they would better reproduce the radiation effects occurring in real life conditions and they would contribute filling a general lack of knowledge. 

In recent times, in-house research facilities at CERN have been occasionally used for material irradiation. In particular the IRRAD facility~\cite{irrad} and the CHARM (CERN High energy Accelerator Mixed field) facility~\cite{charm} are used for proton and mixed field irradiation, respectively. Despite their strategic importance for the testing of electronics and detectors, they have limitations concerning the irradiation of materials. In both IRRAD and CHARM, the available particle fluences do not allow uniform doses in the MGy range to be delivered to macroscopic material samples. For this reason, new research facilities better tailored to the specific requirements of material irradiation are needed.

In the present paper, a new irradiation station recently built at CERN for various applications including material studies is presented. The station, whose installation was finalised in June 2021, allows multiple material samples to be irradiated at different dose rates in a neutron-dominated environment, over exposure times ranging from months to several years. The available radiation fields are comparable to the ones actually present in high radiation areas at CERN, in terms of both dose rate and radiation spectra. The new irradiation station at n\_TOF will therefore meet the twin requirements of producing data that will facilitate the selection of materials for high-radiation areas at CERN and that will contribute to increasing the general scientific understanding in this domain.

To the best of the authors' knowledge, irradiation of liquid or semi-solid samples such as lubricants in mixed fields up to MGy levels, generating sample activation, and requiring containers being compatible with robot handling is not common, at least in a published form. The authors are not aware of any available standard or scientific reference on irradiation set-ups complying with a similar combination of technical constraints, and believe that this is the first time that such a methodology is being systematically addressed.

Section~\ref{sec:near} describes the area at the n\_TOF facility where the irradiation station has been constructed, hereby called "NEAR", along with other areas dedicated to different irradiation and activation experiments. Section~\ref{sec:irrstation} describes in details the technical requirements and the design aspects of the irradiation station and of the containers used for both solid and liquid material samples. Section~\ref{sec:handling} describes the handling procedures necessary for the installation of the irradiation station and for the installation and retrieval of the samples, that must be performed by telemanipulation due to the high activation of the target area. Section~\ref{sec:dosimetry} describes the Monte Carlo calculations performed to assess the level and homogeneity of the neutron and gamma dose absorbed by the samples in different irradiation positions. Section~\ref{sec:RP} describes the radiation protection aspects of the operating procedures of the irradiation station, reporting the residual dose rate in the NEAR area and the calculated activation of samples and containers, that require special handling and disposal as radioactive materials. Section~\ref{sec:firstirradiations} describes the first irradiation of material samples and the first dose measurements carried out between July and November 2021, allowing the whole procedure to be experimentally verified. Finally, conclusions are presented in Section~\ref{sec:conclusions}. 

\section{The n\_TOF facility and its NEAR area}
\label{sec:near}

\subsection{n\_TOF facility: introduction and recent upgrades}

CERN is equipped with a top-class, high-brightness, neutron spallation source dedicated to high-resolution neutron time of flight experiments: the n\_TOF Facility~\cite{ntofweb}. The facility was built in 2000~\cite{borcea} and has largely evolved over the last 20~years. In particular, during CERN’s Long Shutdown~2 (2019-2021), a major upgrade was implemented to guarantee reliable operation of n\_TOF for the following years, improving the physics reach of the infrastructure.

Recent upgrades include the construction of a third-generation spallation target~\cite{esposito2021,esposito2020}, the consolidation of the neutron collimation systems, the complete overhaul of the target pit shielding as well as the construction of a new irradiation station in the NEAR area, close to the neutron spallation target.

\subsection{The NEAR area}
\begin{figure*}[htbp]
\centering 
\includegraphics[width=0.8\textwidth]{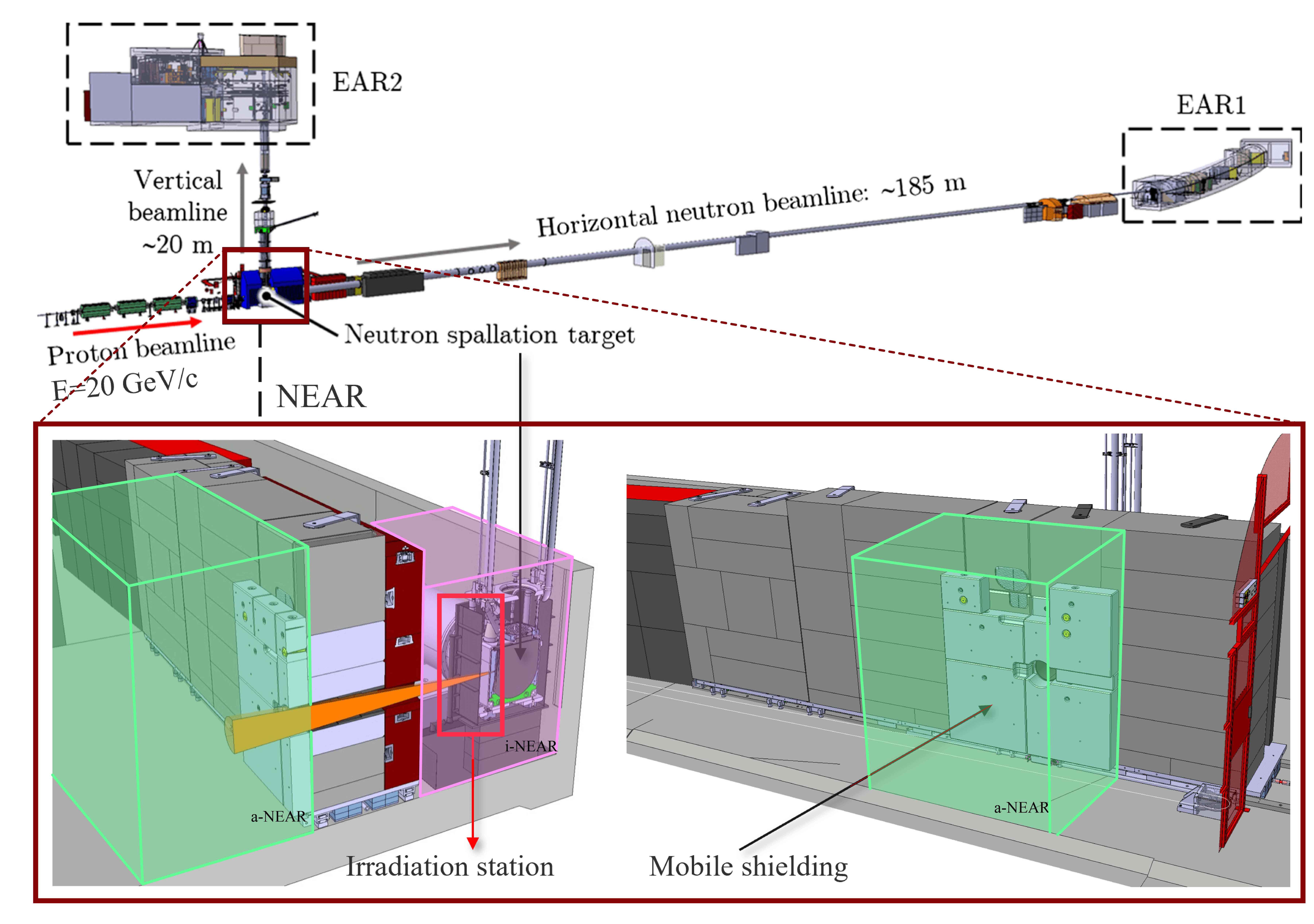}
\caption{\label{fig:NEAR} Overview of the NEAR area within n\_TOF. The upper part of the Figure shows a general scheme of the n\_TOF facility. The proton beamline, the position of the neutron spallation target and the NEAR area (in the brown square) are shown. The lower part of the Figure shows different views of the NEAR area. The position of the irradiation station is indicated by the red rectangle. The mobile shielding, whose position is indicated by the arrow, separates the internal i-NEAR (the pink volume), where the irradiation station is located, from the outer a-NEAR (the green volume).}
\end{figure*}

Figure~\ref{fig:NEAR} shows a schematic overview of the n\_TOF experimental areas. The area close to the new n\_TOF neutron spallation target is referred to as NEAR. The New Target Mobile Shielding (simply the shielding in the paper), separates the NEAR Irradiation area (i-NEAR), which includes the target and where the irradiation station is installed from the NEAR Activation area (a-NEAR), which includes a neutron collimator and a volume outside of the target shielding. 

\begin{figure*}[htbp]
\begin{center}
\subfigure[][]{\includegraphics[width=.28\textwidth]{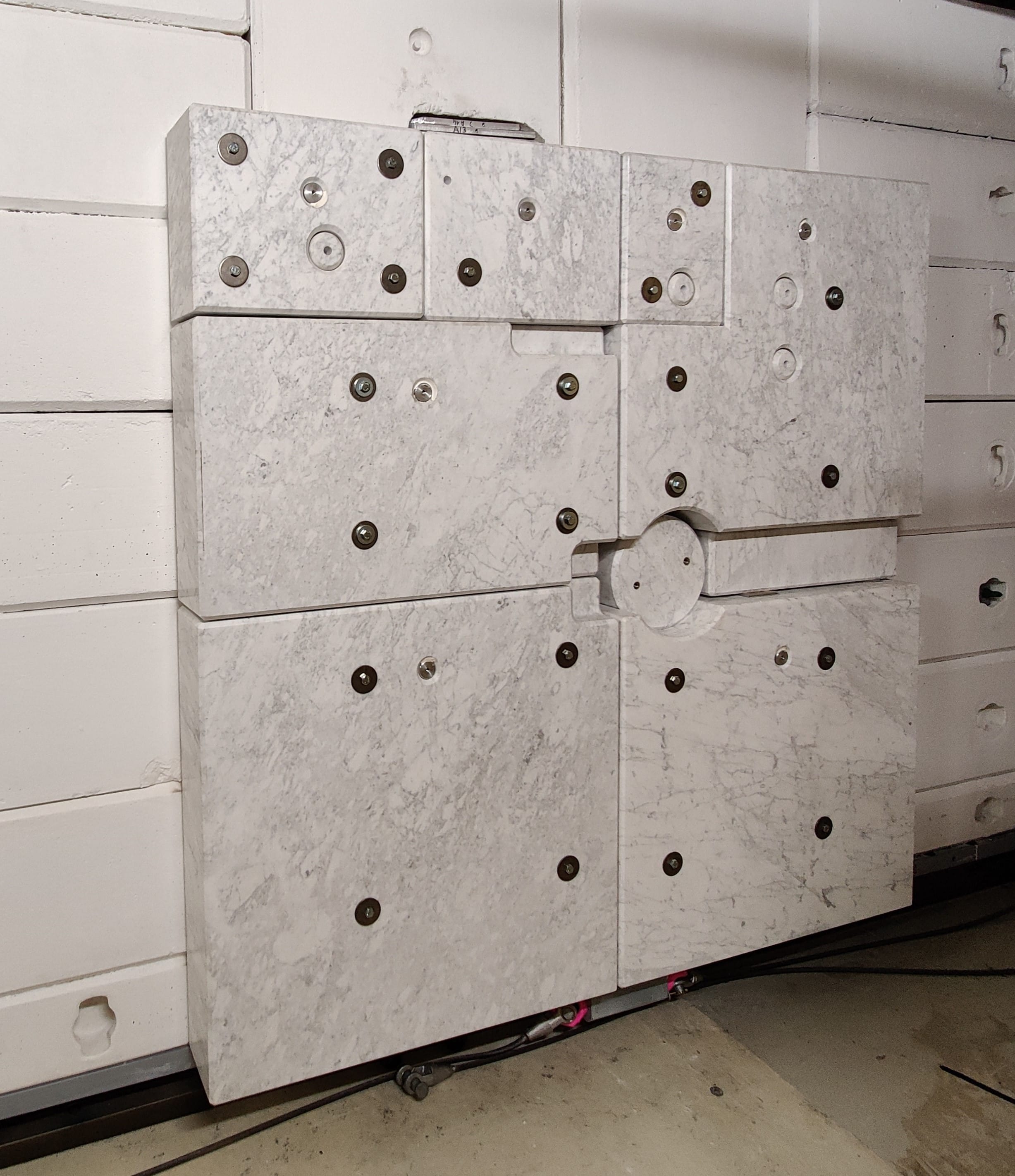}}
\subfigure[][]{\includegraphics[width=.255\textwidth]{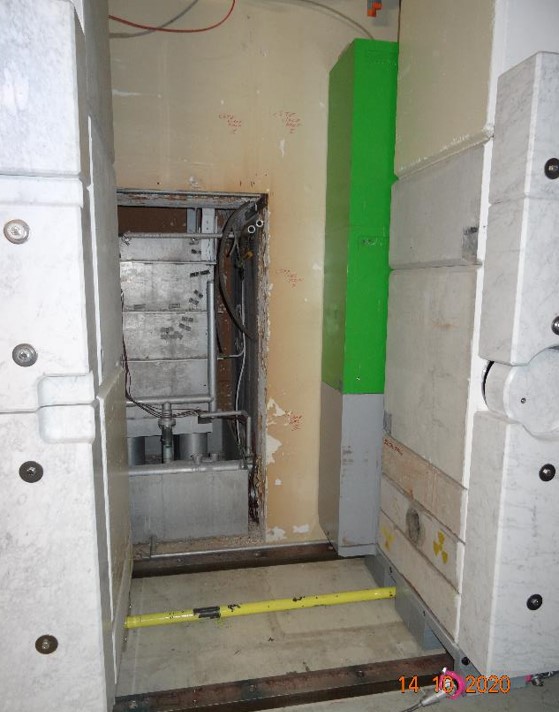}}
\subfigure[][]{\includegraphics[width=0.4\textwidth]{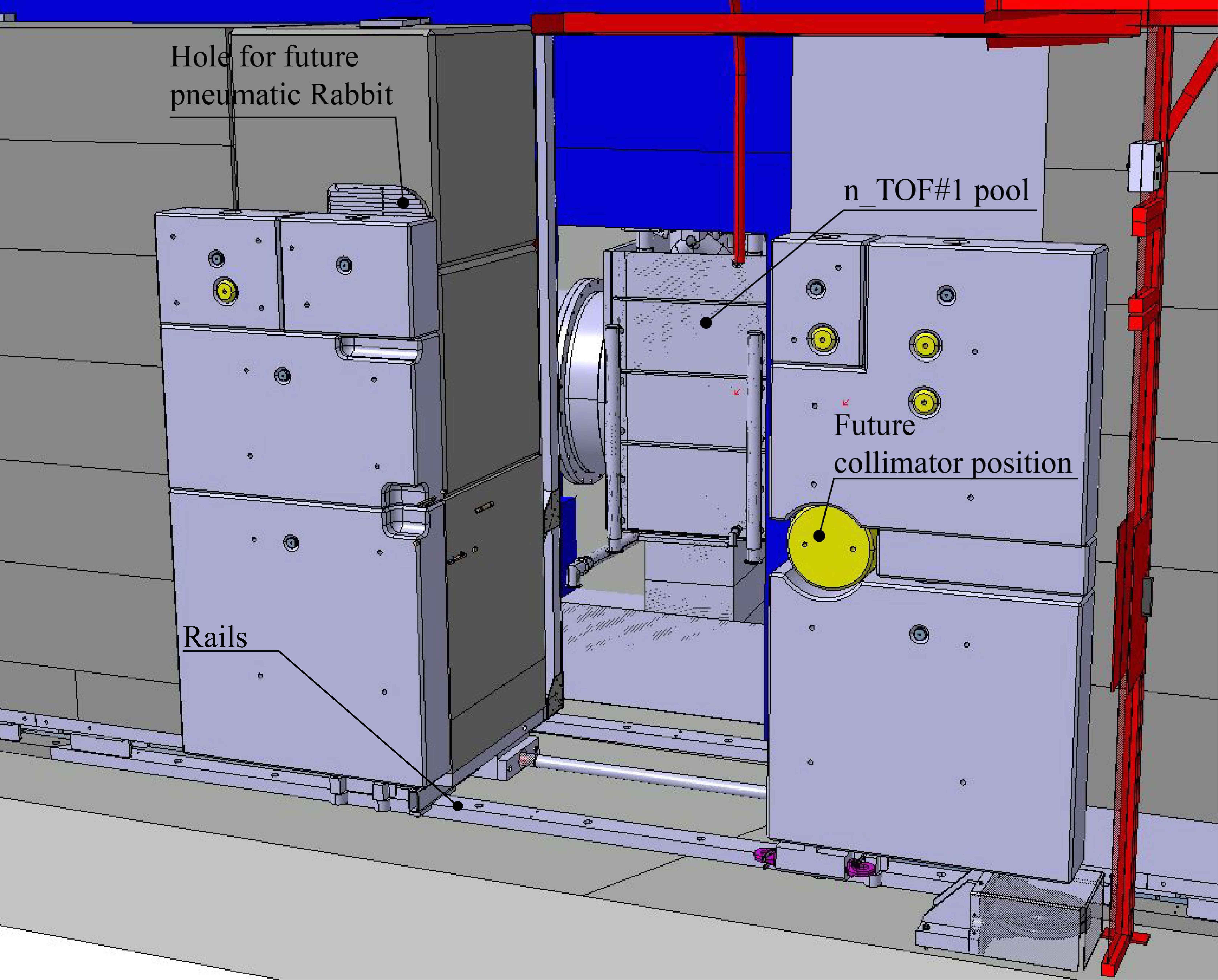}}
\end{center}
\caption{\label{fig:shielding} a) The New Target Mobile Shielding in the closed position. b) The shielding in the open position, revealing i-NEAR and the target pool, close to the n\_TOF target as it was before the installation of the irradiation station. c) 3D model of NEAR, with the  shielding open, showing the position of the target pool, of the future collimator (in yellow), of the openings for rabbit tubes and of the rails allowing the shielding movement.}
\end{figure*}

Figure~\ref{fig:shielding} shows NEAR before the installation of the new irradiation station for materials. As described in Section~\ref{sec:irrstation}, the irradiation station, located in i-NEAR, consists of two shelves (referred to as top and bottom shelf) installed on the side of the existing original target pool where the new third-generation n\_TOF target~\cite{esposito2021} is located. The irradiation station installed in i-NEAR is shown in Figure~\ref{fig:irradstation}. 

The target pool (see Figure \ref{fig:shielding} b and c) is constituted by an aluminium alloy (EN AW-6082) structure which was containing the first generation n\_TOF spallation target~\cite{nTOF:1258424}; the uncladded target was immersed in demineralized water in order to remove the heat generated by the beam interaction with the Pb as well as to moderate neutrons to be employed in the first experimental area. In later generation targets, the pool was left in place in order to act as an external containment in case of water leaks from the cooling or from the moderator circuits.

In addition to the mentioned irradiation station, this highly versatile area includes additional irradiation positions in a-NEAR and channels passing through the shielding, as briefly described in~\ref{sec:rabbit}.

\subsection{The mobile shielding}

\begin{figure}[htbp]
\centering 
\includegraphics[width=0.48\textwidth]{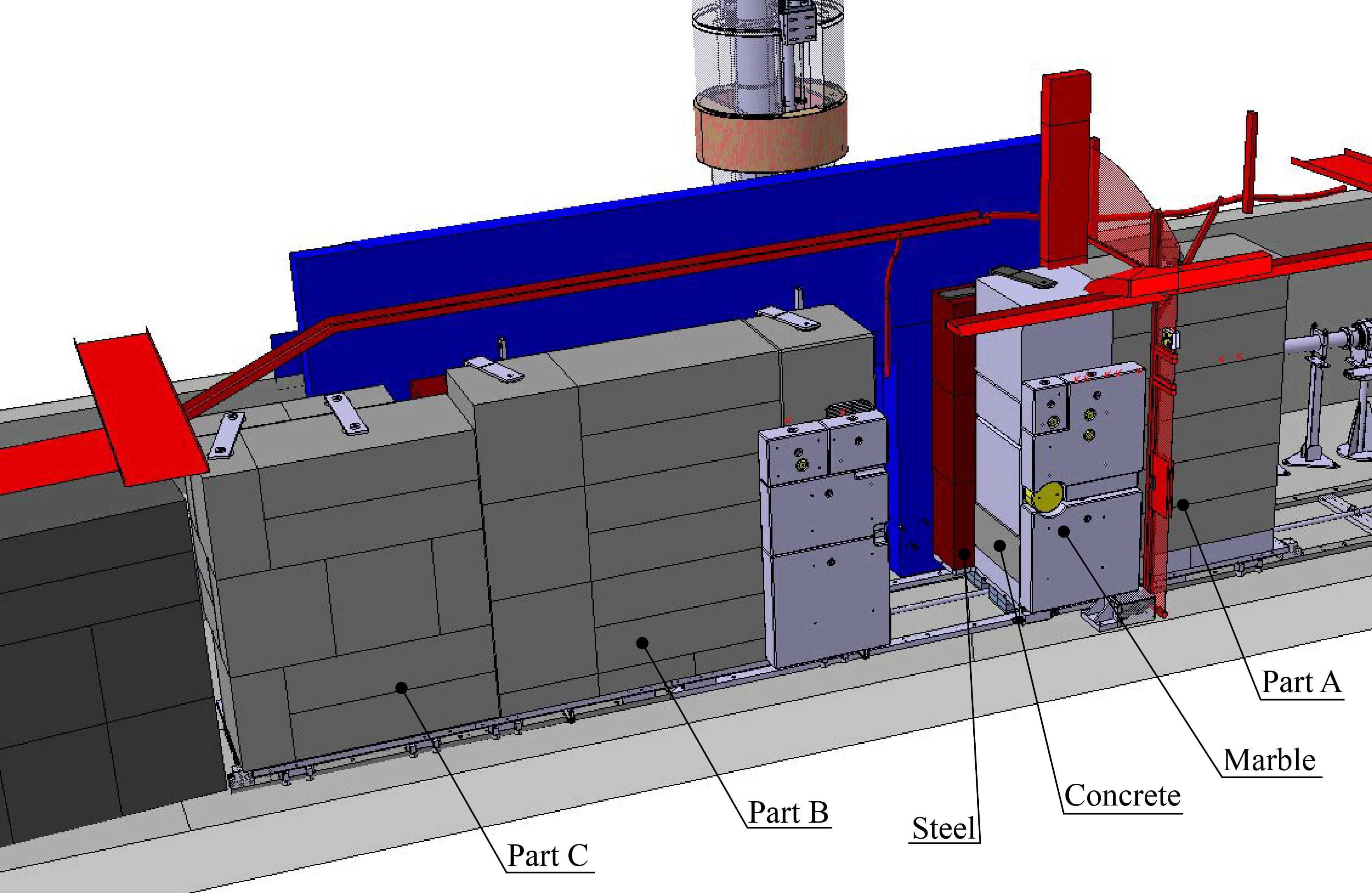}
\caption{\label{fig:shielding_parts} New configuration of the mobile shielding in open position, with part A, B and C, as explained in the text. The different parts and the materials used are indicated. The outermost layer of marble is only present for part A and for part B (for the latter case, only in the area facing the spallation target).}
\end{figure}

The mobile shielding consists of two layers covering the whole length of the target chamber and one partial layer directly in front of the target. The first (inner) layer consists of 400-mm-thick steel (GG20), the second (outer) layer is 800-mm-thick concrete and the third partial layer is 200-mm-thick marble. The shielding is divided in three parts, as presented in Fig.~\ref{fig:shielding_parts}. Part A is the section that is regularly moved to access i-NEAR, while part B is generally never operated, and it is there to guarantee full access to the spallation target in case of need. Part C - only constituted by 1200-mm-thick concrete - is to be moved only for the final dismantling of the facility. The mobile shielding is mounted on top of heavy-duty roller blocks that run on steel rails fixed to the floor. Part A of the mobile shielding is opened and closed using a manually operated winch. The rails allowing the shielding movement are visible in Figure~\ref{fig:shielding}, in the open configuration. The rails are 100~mm wide and 50~mm high, and are an obstacle for the access to the irradiation station by robots, as described in Section~\ref{sec:handling}.

The shielding can be opened during machine stops such as Year-End Technical Stops (YETS), as described in Section~\ref{sec:RP}, to allow access to i-NEAR. Figure~\ref{fig:shielding} shows the shielding in both closed and open positions. With the shielding in the open position, the following tasks can be carried out: access to the target pit, remote inspections of the spallation target, of the moderator and cooling circuits serving the spallation target systems. Moreover, the shielding is designed for manual operation from distance to reduce dose to personnel during interventions and final dismantling.


\subsection{The collimation system}

\begin{figure}[htbp]
\begin{center}
\subfigure[][]{\includegraphics[width=.265\textwidth]{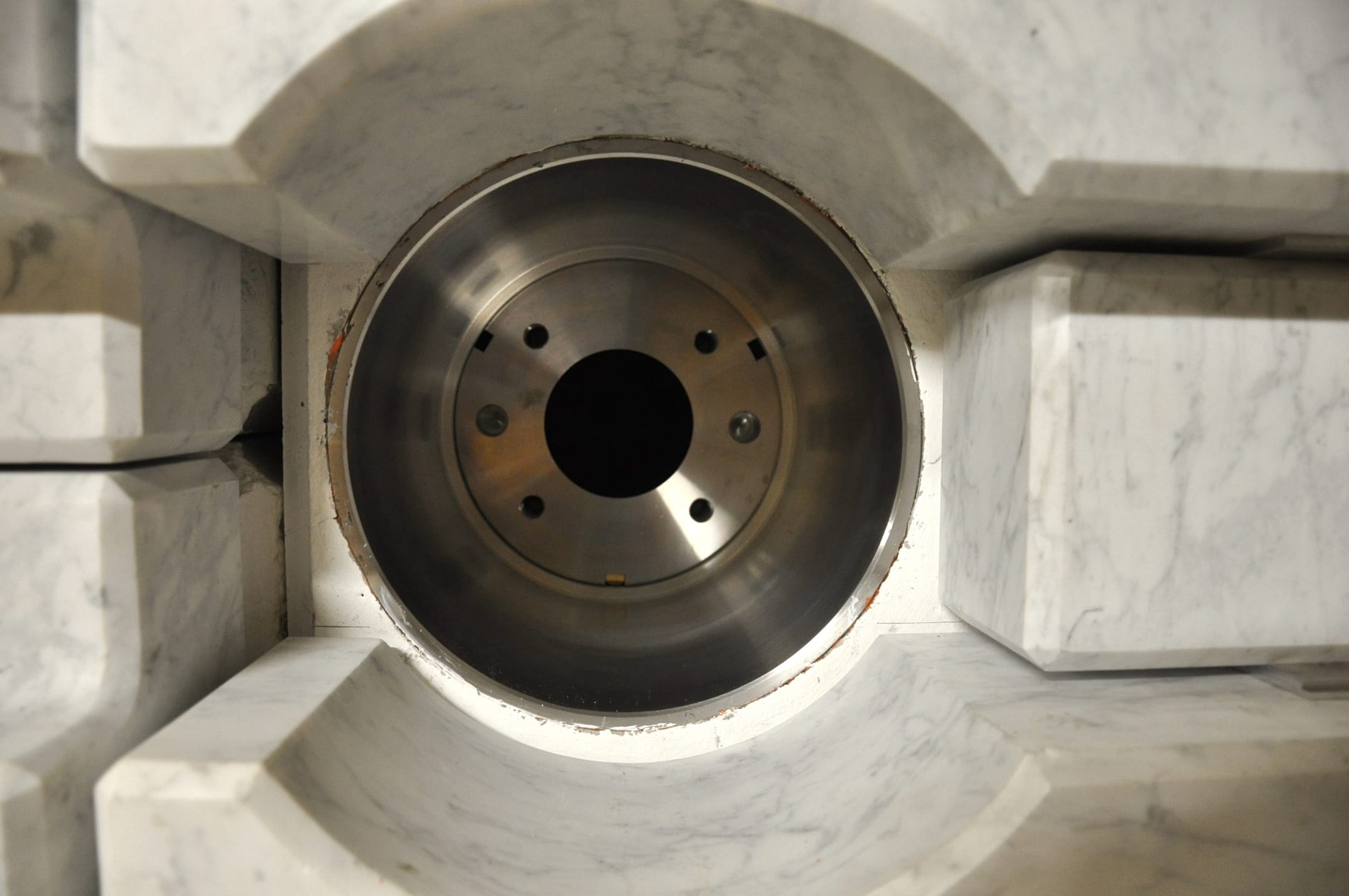}}
\subfigure[][]{\includegraphics[width=.2\textwidth]{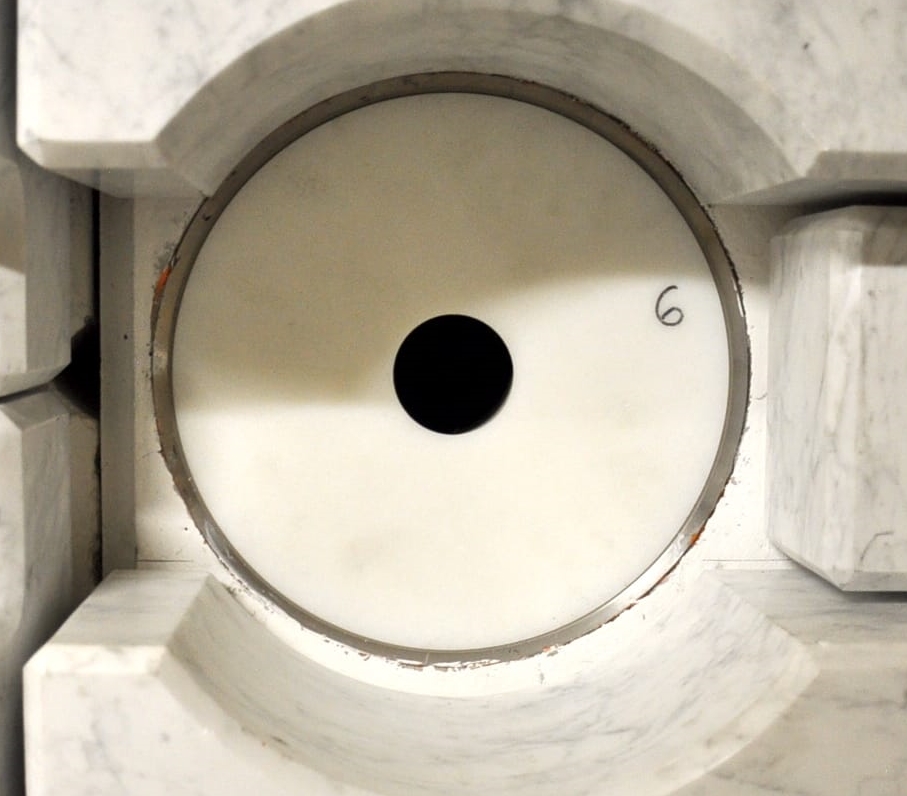}}
\subfigure[][]{\includegraphics[width=0.48\textwidth]{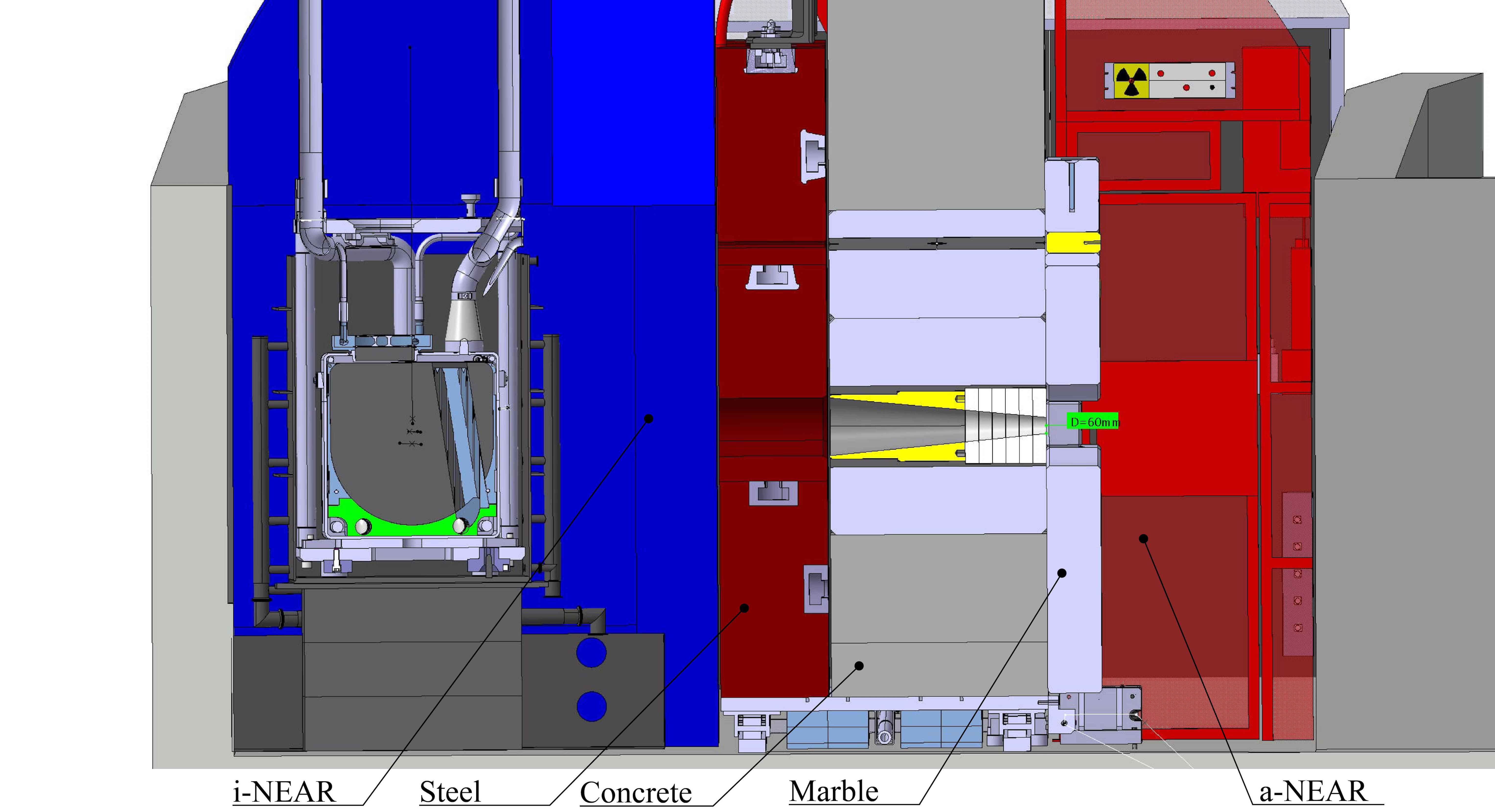}}
\end{center}
\caption{\label{fig:collimator} a) Steel part of the collimator photographed from a-NEAR before installation of the borated polyethylene part. The hole in the marble part of the shielding is visible at the front of the image. b) Borated polyethylene part of the collimator installed in the shielding. c) Section view of the NEAR area showing the target, shielding (steel in red, concrete in grey and marble in light grey) and access passage. The two parts of the collimator are visible inside the shielding: the steel component (in yellow) and the borated polyethlene component (in white).}
\end{figure}

The shielding is designed to permit the installation of a dedicated and flexible collimation system in the proximity of the third generation target~\cite{esposito2021}, in a cylindrical hole of about 280~mm in diameter (see Figure~\ref{fig:shielding}). The collimator (see Figure~\ref{fig:collimator}) consists of two cylindrical layers with a truncated cone-shaped hole in the middle, aiming at giving the desired shape to the radiation coming from i-NEAR and entering a-NEAR. The first (inner) layer consists of a 500-mm-long stainless steel component, the second (outer) layer is made of 300~mm of 5\% borated polyethylene. The configuration has been studied to provide a shaped neutron beam for n\_TOF experiments~\cite{LOI_NEAR}. When not in use, the collimator can be removed and replaced with a concrete (or equivalent) cylindrical structure, to ensure the necessary shielding and to reduce the dose to personnel.

\subsection{Rabbit tubes and other possible irradiation positions}
\label{sec:rabbit}

Apart from the irradiation positions provided by the shelves, other positions are currently being designed and developed to optimise the radiation fields available in both NEAR areas. For example, several channels passing through the shielding, referred to as rabbit tubes, have been reserved to irradiate samples in i-NEAR and allow their retrieval without the opening of the shielding (see Figure~\ref{fig:shielding}). Additionally, specific positions in a-NEAR can be used to explore radiation damage to electronics.

An area dedicated to electronics testing is in fact available on the wall opposite to the neutron collimator, for which a patch panel is available, enabling connection to the control room and therefore the possibility of performing active tests on electronics. An assessment using commercial electronics previously tested in different spallation neutron facilities was performed, confirming that the neutron flux and spectra (see Sections~\ref{sec:spectra} and~\ref{sec:fluence}) are suitable for Single Event Effects testing for atmospheric and high-energy accelerator applications~\cite{cecchetto2022}. 

The present paper concentrates on the description of the design and construction of the shelf positions of the irradiation station in i-NEAR. Further details of the other irradiation positions will be presented in separate publications~\cite{LOI_NEAR}.

\section{Irradiation station for materials: the shelf positions}
\label{sec:irrstation}

\subsection{Opportunities offered by the new irradiation station}

The aim of the irradiation station is to profit from the intense secondary mixed radiation available at NEAR to study radiation damage of materials. Long irradiation exposures ranging from months to years will allow total absorbed doses in the MGy range to be delivered to organic and inorganic materials. As described in detail in Section~\ref{sec:dosimetry}, most of the dose is expected to be delivered by neutrons and to a lesser extent by photons. 

With the NEAR irradiation station, CERN is now equipped with twenty-four irradiation positions, where unprecedented radiation damage data can be collected. In fact, in contrast to standard gamma sources, experimental targets generate unique combinations and spectra of mixed radiation fields, which are hard to replicate in testing environments.

In the irradiation station, a selection of materials used at CERN in high radiation areas will be tested, to further understand the radiation damage mechanisms and to improve their qualification in terms of radiation resistance. The irradiation conditions available at NEAR are much closer to the operational ones when compared to gamma radiation alone. Examples of the irradiation studies to be performed at n\_TOF NEAR are provided in Section~\ref{sec:firstirradiations}.

\subsection{Technical challenges and constraints}

The design and operation of the new irradiation station has to comply with several technical requirements, discussed as follows.

The existing radiation fields available at n\_TOF are produced for the purpose of studying neutron-nucleus interactions, with a broad range of applications~\cite{rubbia}. To minimize any perturbation to the physics performance of the facility, the designs and positions of the new irradiation station structures have been extensively cross-checked with Monte Carlo simulations. 

The use of the irradiation station depends on the n\_TOF facility schedule. The access to the irradiation station will be typically limited to Year-End Technical Stops only, with minimal impact on the operation of the facility and on other physics experiments realized in the same experimental area.

The irradiation station can only be accessed when the shielding is open. As further discussed in Section~\ref{sec:RP}, at around 40~cm from the spallation target residual dose rates typically range between tens and hundreds of mSv/h, after cool-down times ranging between days and weeks. These values are too high for human operators, so access to the irradiation station is generally forbidden. 

Accordingly, the irradiation station was designed to be fully compatible with remotely handled installation. Both the installation of the station structure and the installation and retrieval of the samples are carried out with the use of telemanipulation systems, as described in Section~\ref{sec:handling}. 

The final disposal of the whole irradiation station is taken into account in the design phase as well. To limit the residual activation, aluminium is preferred to stainless steel. The total amount of radioactive waste is limited by opting for a modular structure compatible with potential future modifications. 

The installation of the irradiation station with a telemanipulation system required some unused components still present in the area, such as old cables and cooling pipes, to be removed, and the i-NEAR to be cleaned afterwards. The pipes to be removed were part of the cooling system used for the previous design of the target; the cooling system for the third generation target is completely different and has a new set of cooling pipes. Their removal eased the access to the irradiation station for the robots, as detailed in Section~\ref{sec:handling}. 

\subsection{Design of the irradiation station}
\label{sec:design}

\begin{figure}[htbp]
\begin{center}
\centering \includegraphics[width=.45\textwidth]{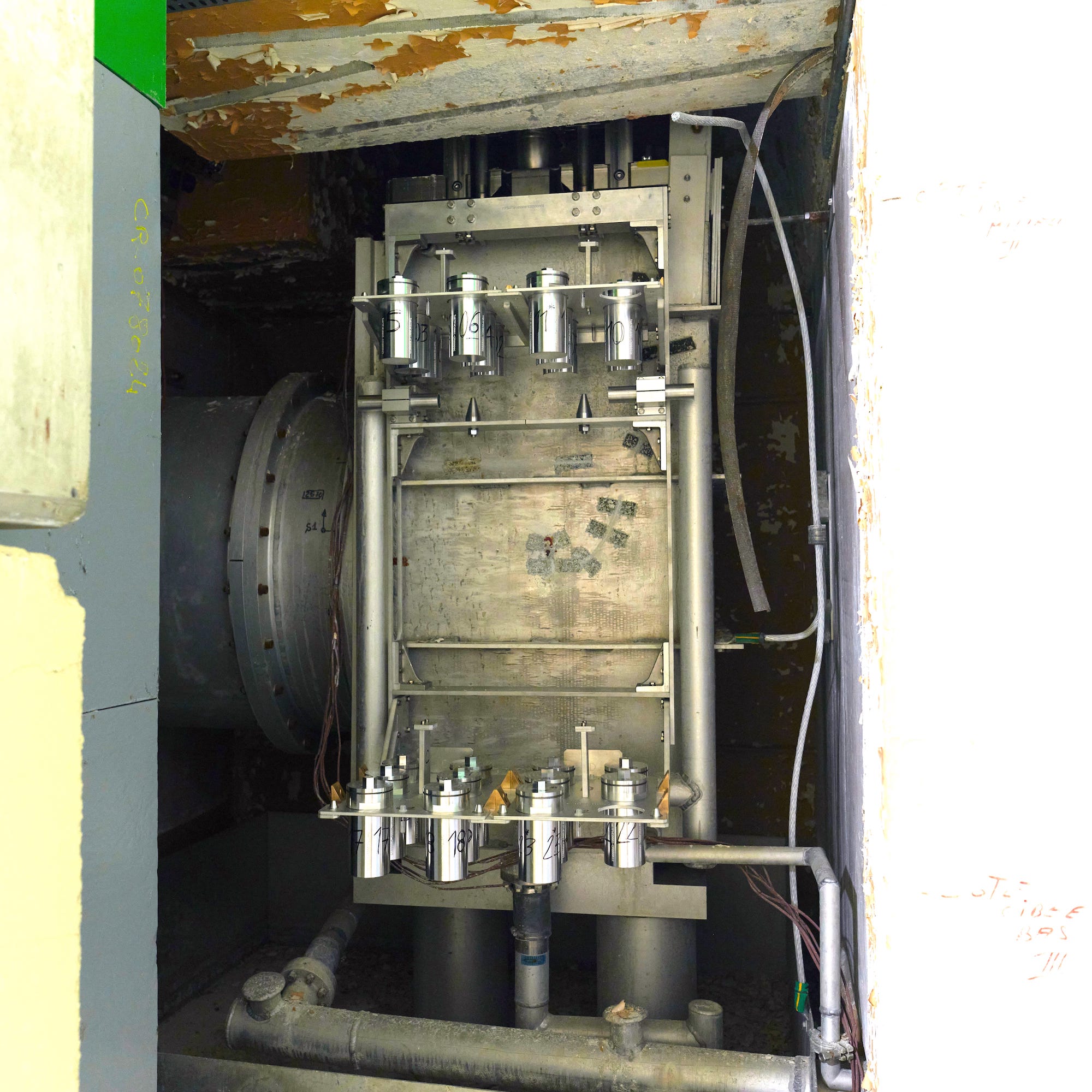}
\end{center}
\caption{\label{fig:irradstation} Picture of the irradiation station, filled with samples, installed on the original target pool at n\_TOF NEAR. The top and the bottom shelves are visible. Four intermediate aluminum supports (two per shelf) with their handle for remote handling, accommodate up to six containers each. (CERN-PHOTO-202107-085-11)~\cite{ordan:2774594}.}
\end{figure}

The new irradiation station is attached to the original target pool by means of hooks at the top of the station frame that engage with the upper edge of the pool wall, which is no longer used for target cooling, as shown in Figure~\ref{fig:irradstation}. In the volume of the shelves, radiation fluences have been calculated to be sufficiently intense and homogeneous for irradiation of materials, based on FLUKA~\cite{Battistoni2015,Bohlen2014,FlukaWeb} Monte Carlo simulations described in Section~\ref{sec:dosimetry}. 

The irradiation station includes two shelves (top and bottom shelf, respectively), as shown in Figure~\ref{fig:irradstation}, and an additional support able to host a dedicated moderator system for a-NEAR, which is outside of the scope of this paper~\cite{LOI_NEAR}. Each shelf can accommodate up to twelve samples, organized and handled in groups of six to facilitate installation and retrieval, when handled remotely as described in Section~\ref{sec:handling}. Up to twenty-four samples can be irradiated at the same time allowing a combination of different exposure times and dose rates, as described in Section~\ref{sec:dosimetry}.

The whole structure is constructed of aluminium EN~AW-6082 T6, has dimensions 61x107.5x57.9~cm and weighs 15.5~kg. Fasteners are made of A4 stainless steel and there are small triangular guides for the installation of the supporting plates made of brass. The structure can safely support a full load of samples and a moderator material block weighing up to 40~kg, to be possibly installed at a later stage. The weight is compatible with the existing original target pool structure, without excessive deformations.

\begin{figure}[htbp]
\centering 
\includegraphics[width=.4\textwidth]{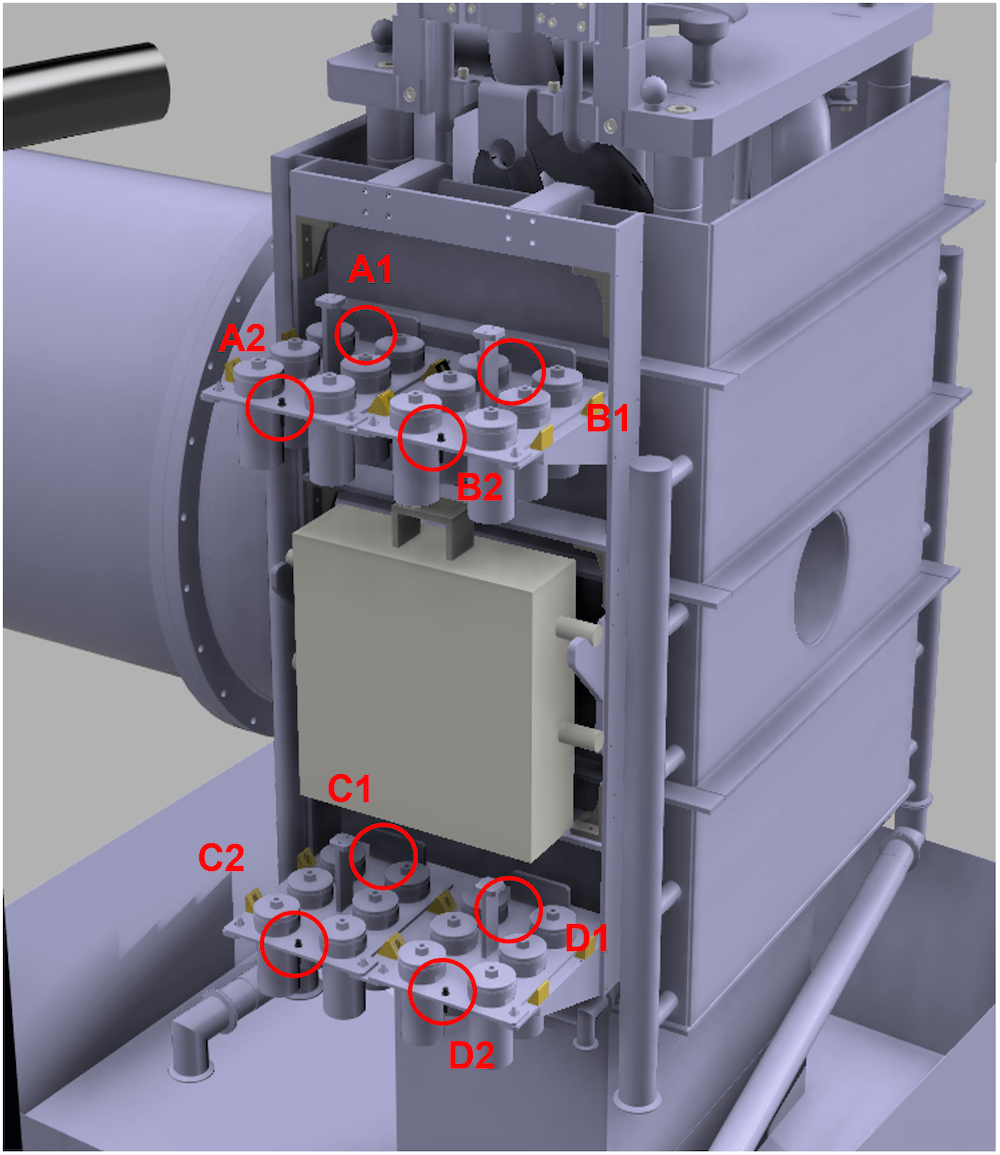}
\qquad
\caption{\label{fig:dosimeter_position} Position of the dosimeters in the irradiation station, along with their labels.}
\end{figure}

Eight locations hosting Radio-Photo-Luminescence (RPL) dosimeters~\cite{vincke2007} (see Figure~\ref{fig:dosimeter_position}) are attached to the shelf structure (four per shelf), allowing the total dose to be mapped during each use of the irradiation station. Readout measurements of the dosimeters irradiated during the first use of the irradiation station are reported in Section~\ref{sec:dosimeters}. The RPL dosimeters are small cylinders made of phosphate glass doped with silver (Ag) ions to improve their luminescence proprieties. When they are irradiated, RPL centers are created within the crystal lattice proportionally to the absorbed dose. The dosimeters were calibrated in a $\textsuperscript{60}$Co gamma-field in the Risø HDRL Facility, Denmark \cite{brynjolfsson1960co} and their measurement range extends up to a maximum of some~MGy of total absorbed dose. The readout measurement is performed at CERN using a custom readout system designed specifically to address the applicability of such dosimeters in high radiation levels \cite{trummer2020characterisation}.

The access to the irradiation station for sample installation and retrieval requires the shielding to be open, and for this reason is only possible during scheduled long technical stops of the n\_TOF facility, such as Year-End Technical Stop.

\subsection{Sample containers: design and use}

Sample life-cycle analysis was used to identify requirements for the design of the containers. The main life-cycle stages to be considered are: container production, robotic or remote installation, retrieval and handling, sample irradiation, post-irradiation analysis and final disposal or decontamination.

The main required features and technical constraints are listed as follows: 

\begin{itemize}
\item Compatibility with a remote handling. Different remote handling techniques are effectively and routinely used to safely deal with activated samples in several facilities, such as research nuclear reactors~\cite{HFIRwebsite}. Hydraulic tube facilities to move shuttle capsules containing the samples are examples of these techniques. In this specific case, the set-up has to be compatible with robot installation; this limits the containers in weight, size and shape. More details are provided in Section~\ref{sec:handling}.
\item Compliance with safety standards; containers must be leak-tight and fall-proof, to reduce contamination risks in case of accidental sample drop during installation or retrieval. A double layer of container is necessary, to confine possible contamination to the disposable inner container only. This allows outer containers to be reused after decontamination, following a cool-down phase. 
\item Radiation tolerance; elastomeric seals should generally be avoided, as even the most radiation tolerant ones are expected to degrade at doses in the MGy range~\cite{yellow3,zenoni2017} (see Section~\ref{sec:dosimetry}). A full-metal container body and graphite seals are therefore used to ensure leak-tightness. 
\item Compatibility with possible pressurisation due to radiation-induced gas production; A stainless steel miniature valve~\cite{valve} with dimensions $\varnothing 8.7\times 10$~mm, a weight of 2.6~g and with an opening pressure of 0.1~bar, is fitted to the container lid, allowing a progressive gas release during irradiation, without compromising their liquid leak-tightness.
\item Optimization for final disposal; aluminium is selected as the main construction material for the containers, to minimize the residual activation originating from neutron irradiation. This is a standard choice for sample irradiation in research reactors~\cite{ferrari2019}. The containers are kept as small and light as possible, to reduce the radioactive waste, in agreement with general radiation protection (RP) principles.
\item Compliance with radiation safety standards; the containers are compatible with handling using custom-made, long-handled 'reachers' and their opening and handling must be easy and quick. This to minimize the radiation dose to the operator, in accordance with RP guidelines. Further details are provided in Section~\ref{sec:RP}.
\end{itemize}

\begin{figure}[htbp]
\begin{center}
\includegraphics[width=.35\textwidth]{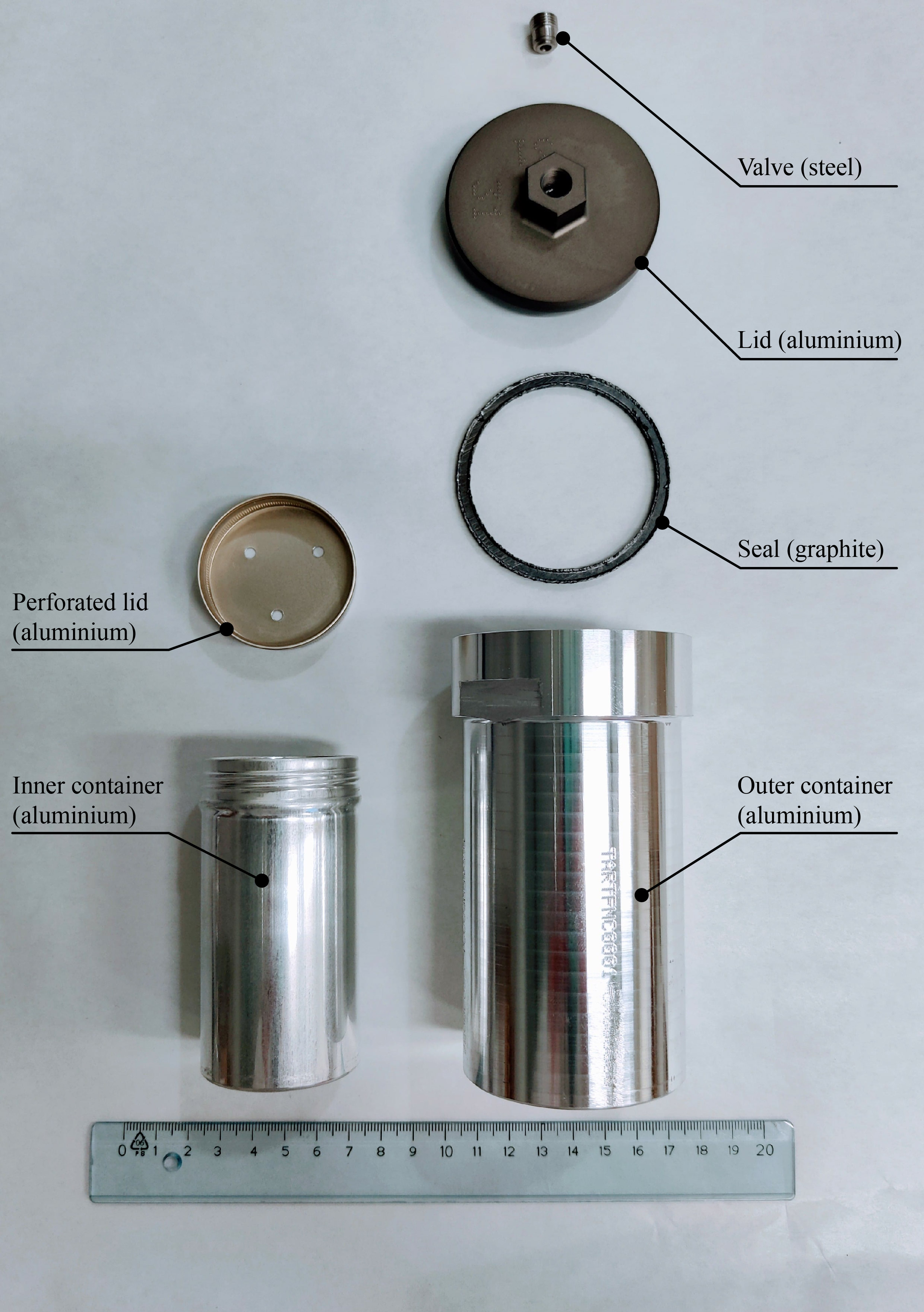}
\end{center}
\caption{\label{fig:container} Sample container design. Left: picture of the aluminium inner container and its perforated lid. Right: picture of the aluminium outer container configuration with its components. From top to bottom: the steel valve, the aluminium lid and the graphite seal.}
\end{figure}

In the final design of the containers, shown in Figure~\ref{fig:container}, all the above-mentioned aspects have been implemented. 

The inner container is a commercially available cylindrical aluminium container with an external diameter of 47~mm and length of 90~mm. A seal of polymeric material is normally glued to the container lid, to ensure leak-tightness. To make the whole configuration radiation tolerant, this polymeric seal has been removed. The lid of the inner container has been pierced to allow the escape of any gas potentially produced during irradiation into the larger outer container. The total inner volume of the container is about 140~cm$^3$. Samples with a volume up to 100~cm$^3$ can be safely hosted by this container. 

The outer container is a custom-made cylindrical aluminium container with an external diameter of 70~mm and length of 142~mm. The graphite seal ensures leak-tightness, while the small valve installed in the lid allows any gas produced inside to escape once the over-pressure reaches the valve opening pressure, without compromising the overall container leak-tightness - for example if the container is dropped during remote handling. The upper part of the lid and of the container body have been shaped to be easily handled, installed and removed from the the irradiation station, and opened. 

\begin{figure}[htbp]
\begin{center}
\includegraphics[width=.45\textwidth]{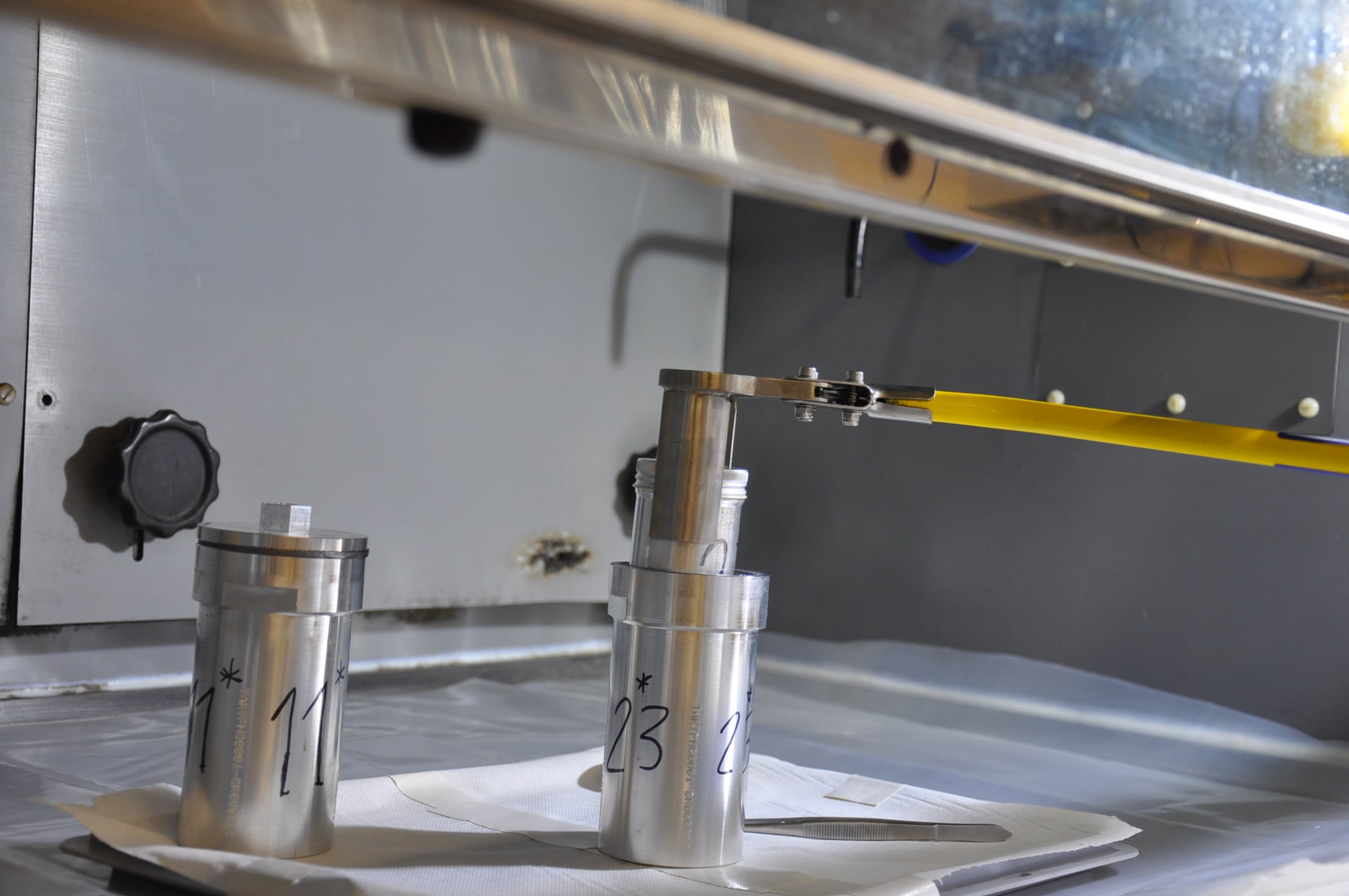}
\end{center}
\caption{\label{fig:pilot} Extraction of an inner container from an outer container under fume hood, by use of the reachers.}
\end{figure}

The inner container can be inserted and extracted from the outer container using custom-designed long-handled reachers (as shown in Figure~\ref{fig:pilot}). The reachers enable the extraction of the inner container without spilling non solid samples and additionally allow an overall distance of 40~cm to be maintained between the operator and the containers. This might be necessary during the post-irradiation handling, depending on the residual activation of the containers and samples.

The samples are installed in the irradiation station after first installing them in a intermediate support. There are a total of four intermediate supports accommodating up to six containers each. An intermediate support consists of an aluminium plate with six circular holes, allowing the sample containers to hang from the plate. Each intermediate support has a handle for remote handling, as described in Section~\ref{sec:installation_samples}. 

The intermediate support and its containers have been designed for ease of installation and removal from the irradiation station shelves. The containers have been designed and tested to ensure that samples will not spill out if they are dropped during handling.

\section{Installation and handling via telemanipulation systems}
\label{sec:handling}

 Manual insertion and retrieval of the samples from the shelves is not possible due to the expected high residual activity of the n\_TOF NEAR area, as described in Section~\ref{sec:RP_NEAR}. Accordingly, the installation of the irradiation station and the installation and retrieval of the samples are performed by a combination of robots~\cite{Teodor, Telemax, di2017dual} operating in front of the target area at NEAR, as shown in Figure~\ref{fig:robot_tasks}.

\subsection{Preparation of i-NEAR}
i-NEAR had to be prepared to host the irradiation station. The robots used needed a combination of strength, precision and specific abilities to remove the old cabling surrounding the target area, as well as the old cooling pipes installed in the pit. The tasks and the robots used to perform them are listed below:

\begin{itemize}
    \item Remove old unused cables from the area. To cut cables, dedicated shears are installed on CERNBot~2.0~\cite{di2017dual, vanden2019robotic}. Teodor \cite{Teodor}, a remote operated vehicle manufactured by Telerob company~\cite{Telerob} and CERNBot~2.0 were used in combination to respectively hold and cut the cables (See Figure~\ref{fig:robot_tasks}-A).
    \item Vacuum cleaning of dirt particles from the pit retention vessel (located below the target pool) and the shielding floor area performed using Teodor.
    \item Dismantling of the cooling pipes used for a previous target design and no longer needed. It was necessary to loosen the clamps attaching the pipes to the pit. This action required a combined used of different robots: whereas Telemax~\cite{Telemax} (a robot manufactured by Telerob company) used an angular screwdriver to reach the clamps, Teodor held the pipes to prevent them from falling on Telemax, as well as to avoid creating undesired debris.
\end{itemize}

After preparation of the area, the irradiation station was installed, as described in the following Section.

\subsection{Installation of the shelves}

All the robots entering i-NEAR have to be able to pass over the floor rails allowing the opening and closing of the shielding. 

\begin{figure}[htpb]
\centering\includegraphics[width=1\linewidth]{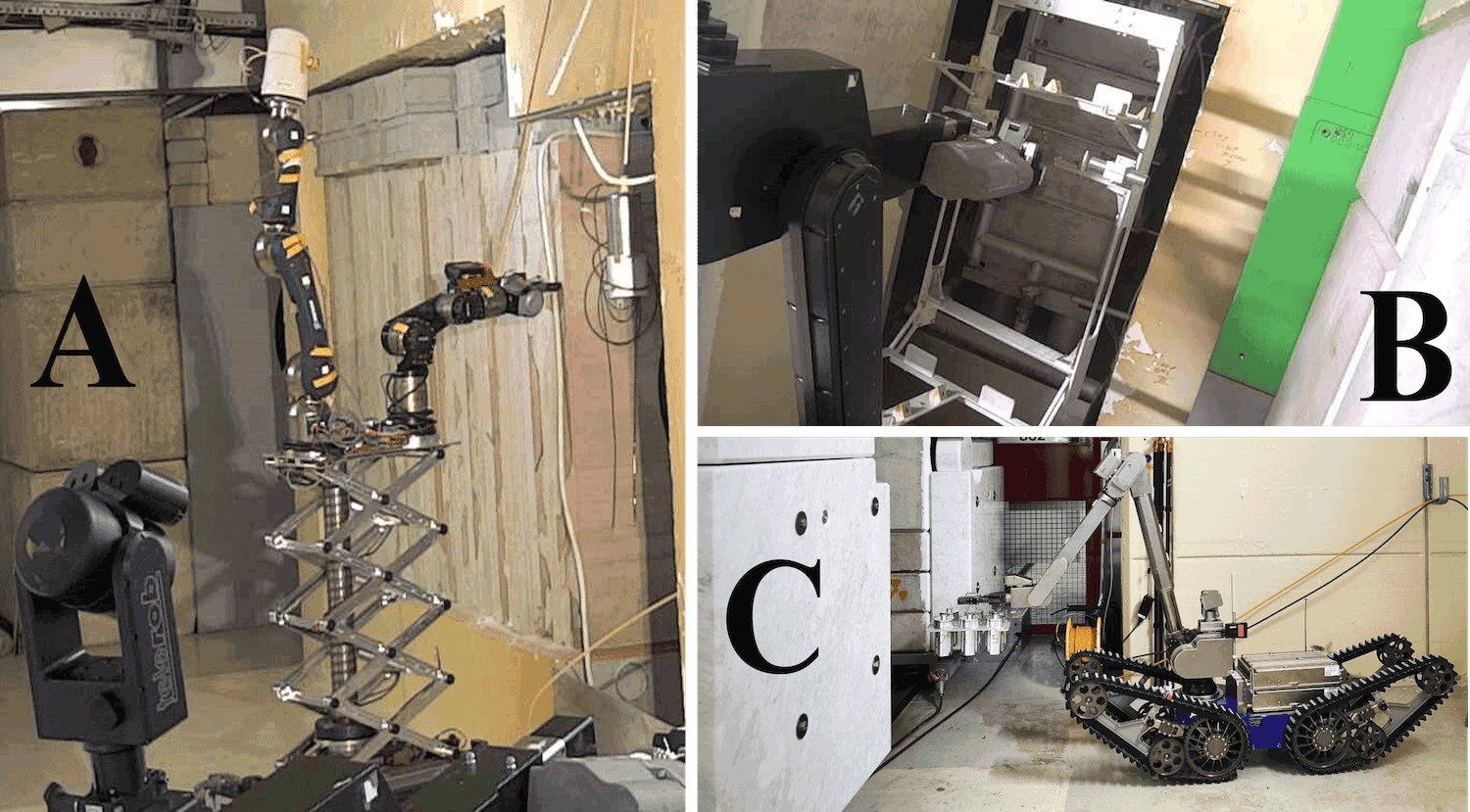}
\caption{Three of the main tasks carried out by robots: A) CERNBot2.0 with double arm configuration cutting cables; B) Teodor installing the shelf structure on the n\_TOF's spallation original target pool; C) Telemax robot installing an intermediate support with 6 sample containers in the NEAR irradiation station (CERN-PHOTO-202107-085-7)~\cite{ordan:2774594}.}
\label{fig:robot_tasks}
\end{figure}

The structure of the shelves has been designed to be compatible with robots: their center of gravity and the area's dimensional constraints were carefully taken into account to meet this requirement. The structure was installed directly on the original target pool. The use of the Teodor robot, having a range of payload up to 100~kg depending on the arm stretch, was required for their installation, as shown in Figure~\ref{fig:robot_tasks}-B. 

The handling of the shelves requires a combination of high precision, and the ability to carry and handle a load of 15.5~kg, corresponding to the total shelf weight. 

\subsection{Installation of the samples}
\label{sec:installation_samples}

\begin{figure}[htpb]
\centering\includegraphics[width=0.9\linewidth]{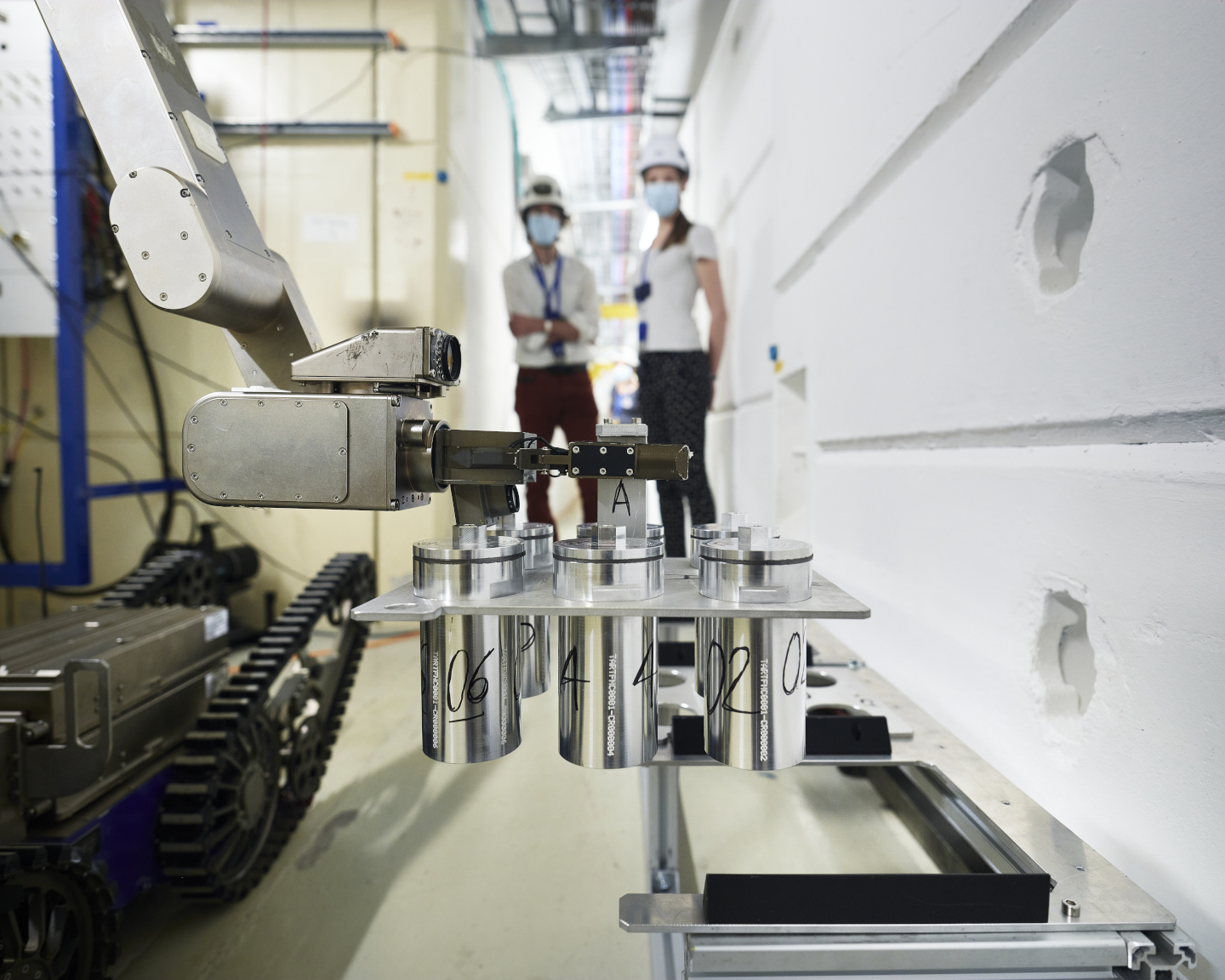}
\caption{Telemax holding a support structure containing six samples of materials for installation (CERN-PHOTO-202107-085-1) \cite{ordan:2774594}.}
\label{fig:sample_handling}
\end{figure}

As shown in Figure~\ref{fig:sample_handling}, the samples are remotely handled in groups of six by use of an intermediate support. The total weight of the intermediate support with the containers and the samples is about 3.6~kg, making it compatible with the use of Telemax, whose lifting capacity ranges up to 20~kg.

The installation of the samples in the shelves is a delicate and challenging step of the handling procedures. In fact, the samples are unavoidably shaken by the movement of the wheel tracks and by the difficult passage of the robot over the shielding rails, necessary to access the target pit. As some of the samples might be semi-solid or fluid (see  Section~\ref{sec:firstirradiations}), they should be kept as much as possible in a steady vertical position. In addition, the narrow available space in i-NEAR adds complication and limitations to the robot movement and degrees of freedom. As shown in Figure~\ref{fig:robot_tasks}-C and in Figure~\ref{fig:sample_handling}, Telemax is chosen to complete the installation and retrieval of the intermediate supports containing the samples of materials to be irradiated, due to its higher dexterity and accuracy.

\subsection{Recovery scenarios}

Given the technical difficulties related to the installation and retrieval of the samples, a complete risk analysis was performed combined with extensive tests of the most likely failure scenarios. A mock-up of the system was used to simulate the potential failure scenarios and training runs were made to prepare for the installation. As a result of this work, the designs of some parts (mainly alignment guides and holding components) were modified to be more robot-friendly and making it is easier for the operator to approach the shelf with a new group of samples and to precisely place each sample group carefully in the irradiation station.

\begin{figure}[htpb]
\centering\includegraphics[width=1\linewidth]{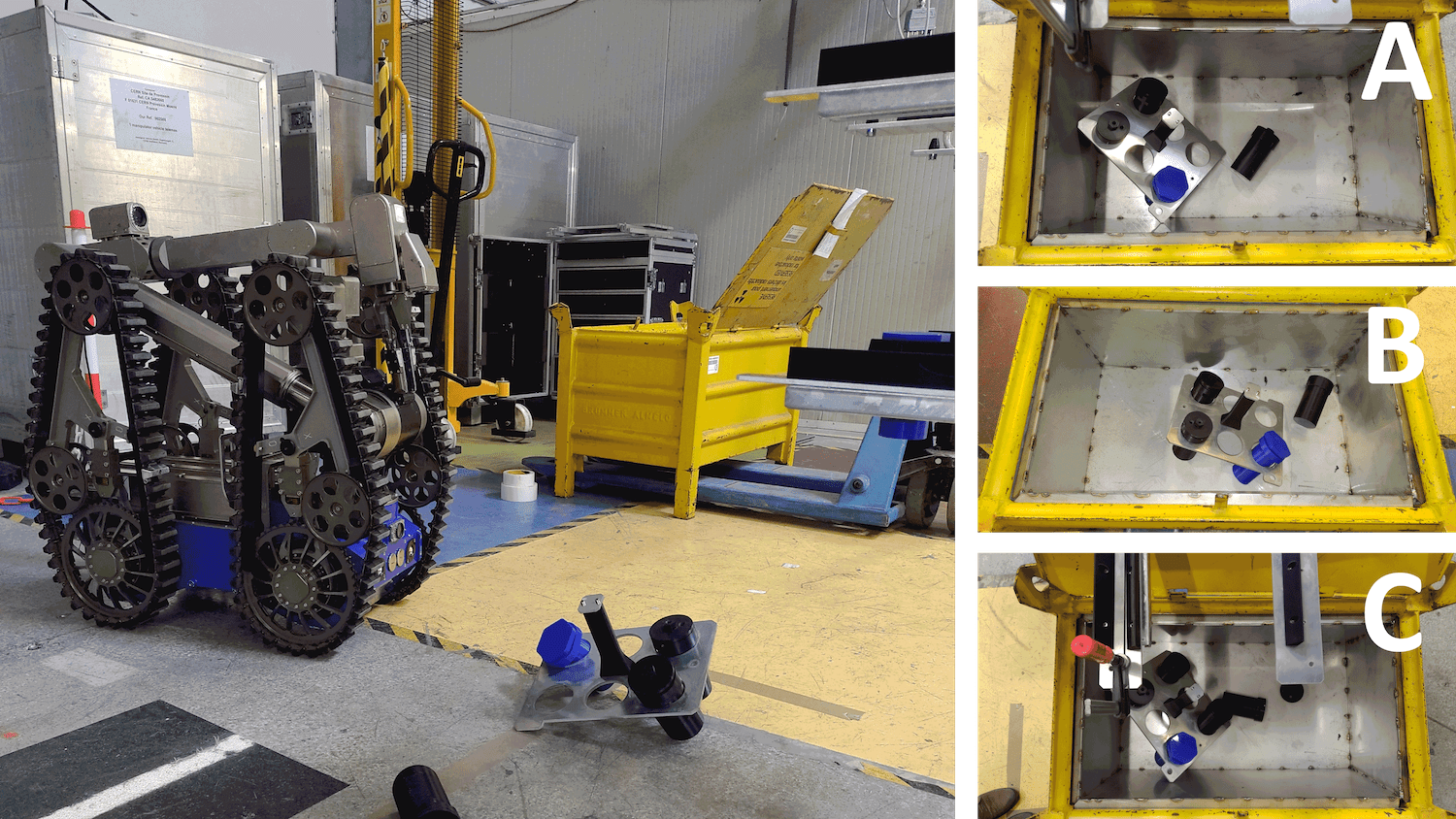}
\caption{Mock-up trials of recovery scenarios: In the left picture Telemax is recovering some dropped containers; an intermediate support is on the floor with a sample having fallen out of it. Right side, simulation of three  possible scenarios (A, B, C) of a dropped intermediate support. The yellow container was used to replicate the retention vessel in i-NEAR.}
\label{fig:recovery}
\end{figure}

This was necessary before the go-ahead was given to start work in the active area. Figure~\ref{fig:recovery} shows recovery trials after simulating a dropped intermediate support and its containers.

In conclusion, the set of available robots, combined with the specific training of the operators, allowed the irradiation station to be successfully installed and safe methodologies to be defined for the installation and retrieval of the samples. 

\section{Dosimetry calculations}
\label{sec:dosimetry}

FLUKA Monte Carlo software~\cite{Battistoni2015,Bohlen2014,FlukaWeb} simulations were used to estimate the total dose absorbed by the samples in the irradiation positions provided by the shelves. The scope of the simulations also included verification that the desired doses can be delivered with sufficient homogeneity and within time intervals compatible with the facility schedule.

\subsection{Geometry and sample composition}

\begin{figure}[htbp]
\begin{center}
\includegraphics[width=1\linewidth]{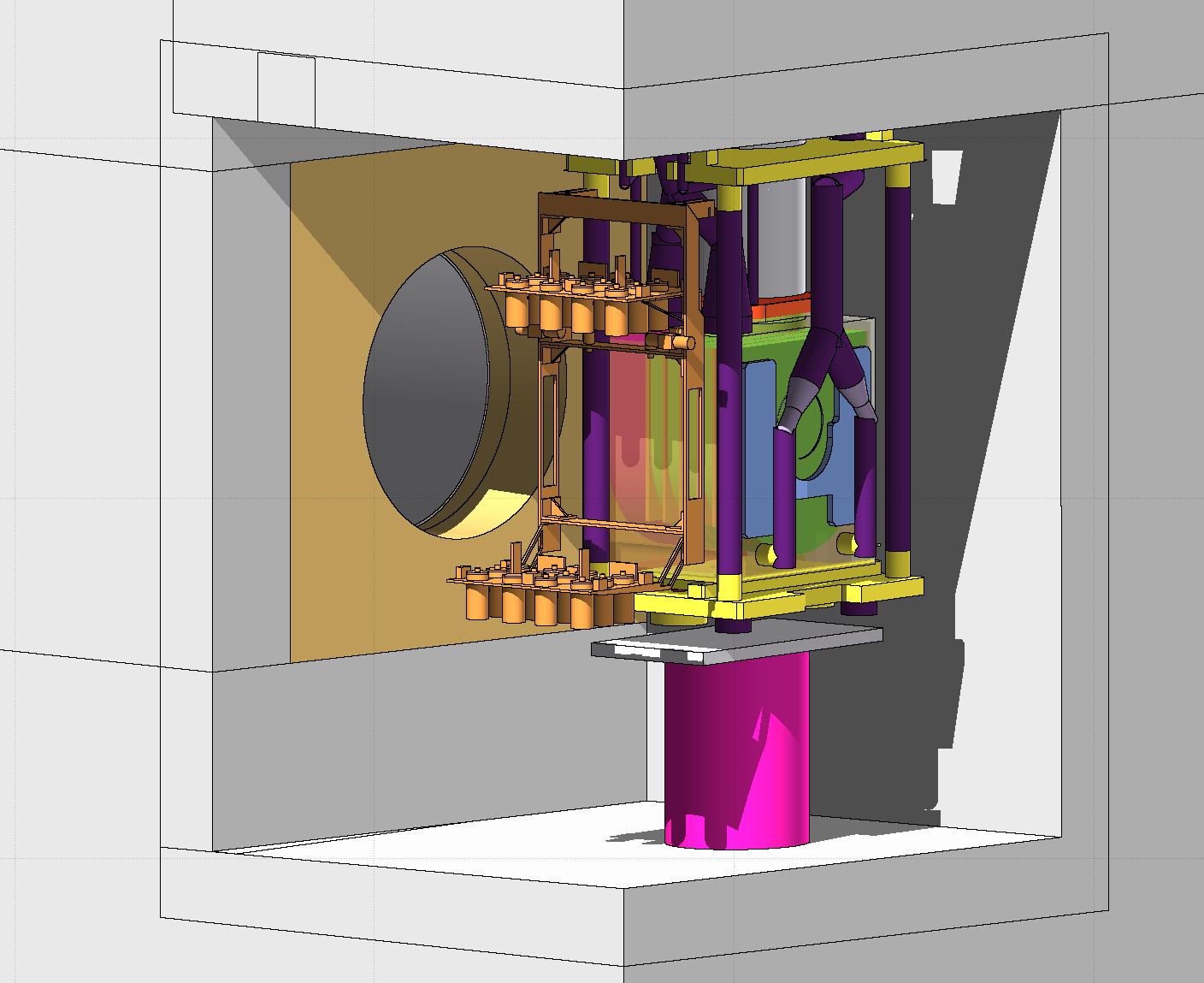}
\end{center}
\caption{\label{fig:flukageo} FLUKA Monte Carlo geometry of the area surrounding the n\_TOF third generation spallation target, focusing on the shelves of the irradiation station.}
\end{figure}

A simplified geometry of the shelves, of the irradiation containers and of the samples has been implemented in FLUKA to simulate the radiation quantities of interest, as shown in Figure~\ref{fig:flukageo}. In the simulation, the irradiation set-up is represented by a cylindrical container made of aluminium having outer diameter of 6.3~cm, inner diameter of 5.6~cm and height of 13.1~cm. The cylinder is generally filled with about 100~$cm^3$ of sample.

Table~\ref{tab:chemcomp} shows the assumed sample composition employed for the Monte Carlo dose assessment. As described in Section~\ref{sec:firstirradiations}, the irradiation station is used to irradiate a wide selection of commercial materials. As a first approximation and based on previous studies~\cite{ferrari2019,ferrari2021oil}, the selected composition can be considered as generally representative of polymeric materials mostly composed of light elements, such as lubricants and elastomeric O-rings. The chosen materials are mostly composed of carbon, hydrogen and oxygen, but metallic traces are usually present in uncontrolled amounts. As described in Section~\ref{sec:RP}, these traces are included to simulate the residual activation of the samples after neutron irradiation, which is mainly due to metals.

\begin{table}[htbp]
\centering
\caption{\label{tab:chemcomp} Chemical composition of the polymeric material used for dose simulations.}
\begin{tabular}{rlrlrl}
\hline
Element & Mass (\%)  \\
\hline \hline 
C & 83.9    \\
H & 10.0  \\
O & 4.3   \\ 
Na & 1.6   \\
Zn & 0.177  \\
Ca & 3.64 $\cdot 10^{-2}$ &  \\
Cd & 3.83 $\cdot 10^{-4}$ \\
Br & 6.87 $\cdot 10^{-5}$ \\
Cr & 2.95 $\cdot 10^{-5}$ \\
La & 2.95 $\cdot 10^{-5}$ \\
Ce & 2.95 $\cdot 10^{-5}$ \\
Sb & 6.87 $\cdot 10^{-6}$ \\
As & 4.91 $\cdot 10^{-6}$ \\
Sm & 3.93 $\cdot 10^{-6}$ \\
U-238 & 3.90 $\cdot 10^{-6}$ \\
Sc & 1.96 $\cdot 10^{-6}$ \\
U-235 & 2.83 $\cdot 10^{-8}$ \\

\hline \hline 
\end{tabular}
\end{table}

\subsection{Absorbed dose}

In the twenty-four irradiation positions provided by the shelves, there is a mixed radiation field dominated by neutrons and photons. The main quantity selected to characterize the dosimetry of the irradiated positions is the physical absorbed dose in a reference sample along with its main contributions, the neutron dose and the gamma dose components. The gamma dose is computed as an electromagnetic dose which includes the contributions from photons, electrons and positrons. Since this contribution is in this case dominated by the gamma component, in the paper it will be simply referred to as gamma dose. All the other particles are neglected, as their contribution is several orders of magnitude lower.

Dose calculations are performed using a proton beam of energy~20~GeV/c \cite{esposito2021,guerrero2013} impinging on the spallation target. It is convenient to express dose values in Gy/pulse, a nominal pulse corresponding to 7$\cdot~10^{12}$ protons \cite{esposito2021,barbagallo2013}. As a general reference based on previous experience, a total of approximately 2.5$\cdot~10^{19}$ protons on target (POT) are expected to be delivered in 200~working days during a normal physics production year, which corresponds to about 1~year of standard operation of the n\_TOF facility. 

\begin{figure}[htbp]
\begin{center}
\subfigure[][]{\includegraphics[width=.45\textwidth]{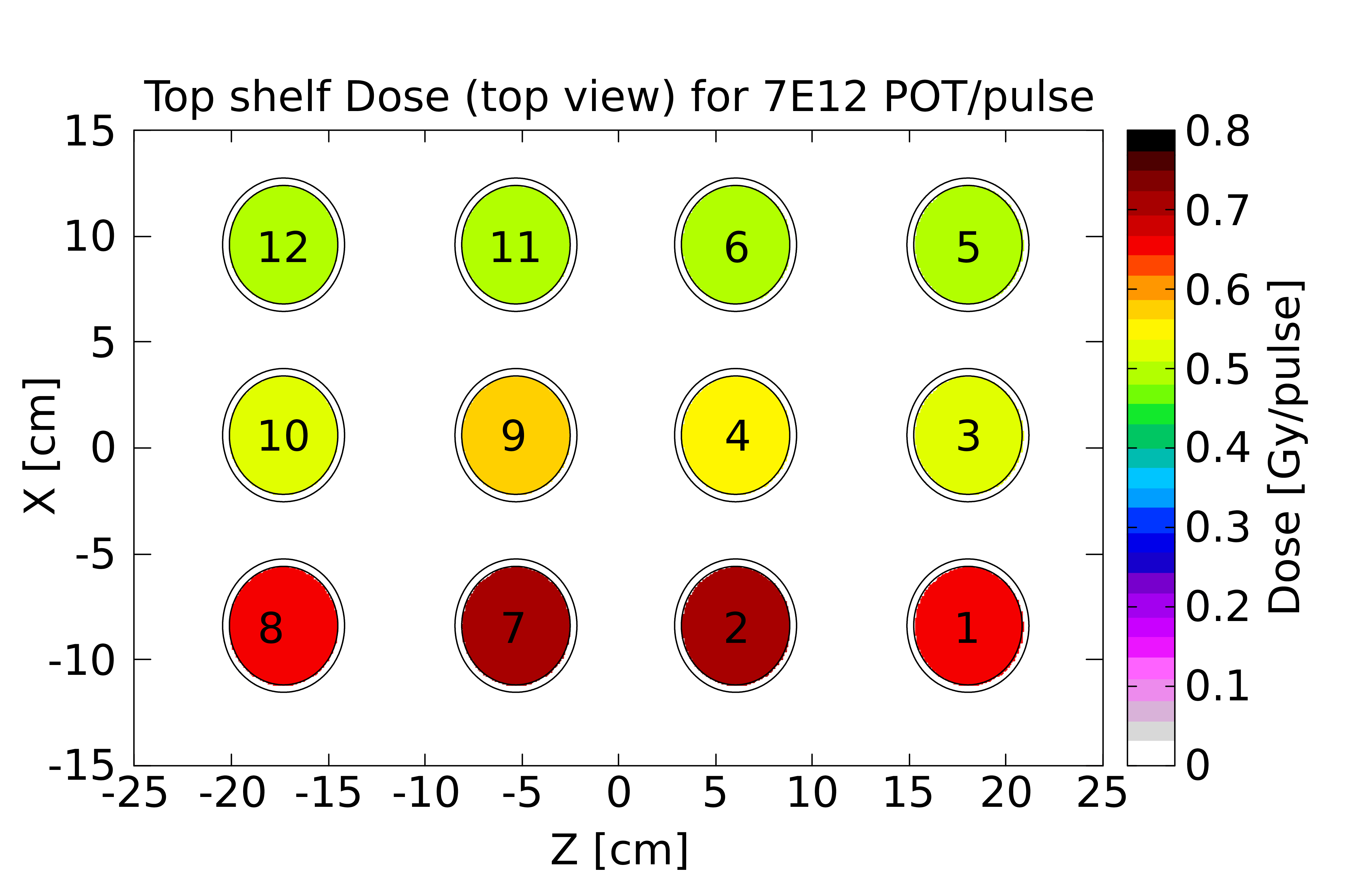}}
\subfigure[][]{\includegraphics[width=.45\textwidth]{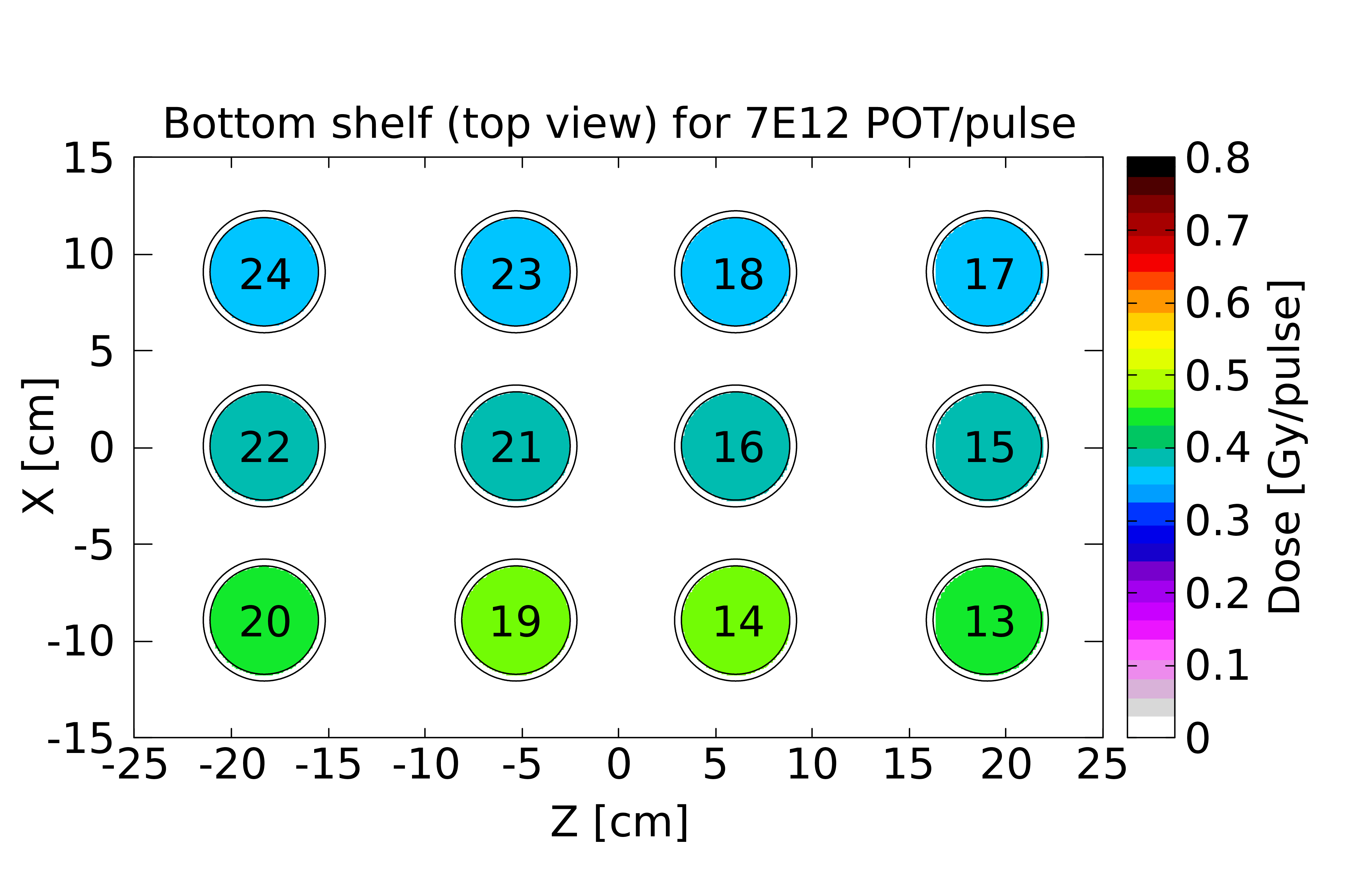}}
\end{center}
\caption{\label{fig:shelf_dose} Total absorbed dose in the samples in the top shelf (a) and in the bottom shelf (b), averaged over each sample volume. Dose values are expressed in Gy/pulse, for a reference value of 2.5$\cdot~10^{19}$ protons on target (POT), expected to be delivered during a normal physics production year of 200~working days. The numbering of the samples corresponds to the different groups of 6 samples handled by the robots. The target, whose position is not shown in the Figure, is further down in x direction, meaning that samples 1-2-7-8 and 13-14-19-20 are the closest to the target (on the top shelf and bottom shelf, respectively).}
\end{figure}

Figure~\ref{fig:shelf_dose} shows the dose absorbed in the samples, averaged over the whole sample volume. On the top shelf, the total doses range from 0.5~Gy/pulse to 0.7~Gy/pulse, due to different distances from the target (x axis in the Figure). On the bottom shelf, the total doses range from 0.35~Gy/pulse to 0.45~Gy/pulse. In a standard physics operation year, the delivered doses correspondingly range from 1.2~MGy (lowest value in the bottom shelf) to 2.5~MGy (highest value in the top shelf).

In both shelves, the samples located at the same distance from the target (such as samples in position 1, 2, 7 and 8 in Figure~\ref{fig:shelf_dose}, and so on) absorb a comparable dose. By contrast, the total dose progressively decreases the farther the samples are laterally with respect to the target, as reported in Table~\ref{tab:doses}. In the Table, dose values are averaged over the four samples located at the same lateral distance from the target. 

The gamma component of the dose is roughly independent of the specific irradiation position: it corresponds to 0.13~Gy/pulse in all the positions of the top shelf, while it corresponds to 0.11~Gy/pulse in all the positions of the bottom shelf. By contrast, the neutron component of the dose ranges from 67\% to 81\% of the total, progressively decreasing with distance from the target. This can be explained considering that organic materials act as moderators of high-energy neutrons, due to the high amount of hydrogen present in their composition. The samples located further from the target are partially shielded by the presence of the closer samples.

\begin{figure}[htbp]
\begin{center}
\subfigure[][]{\includegraphics[width=.45\textwidth]{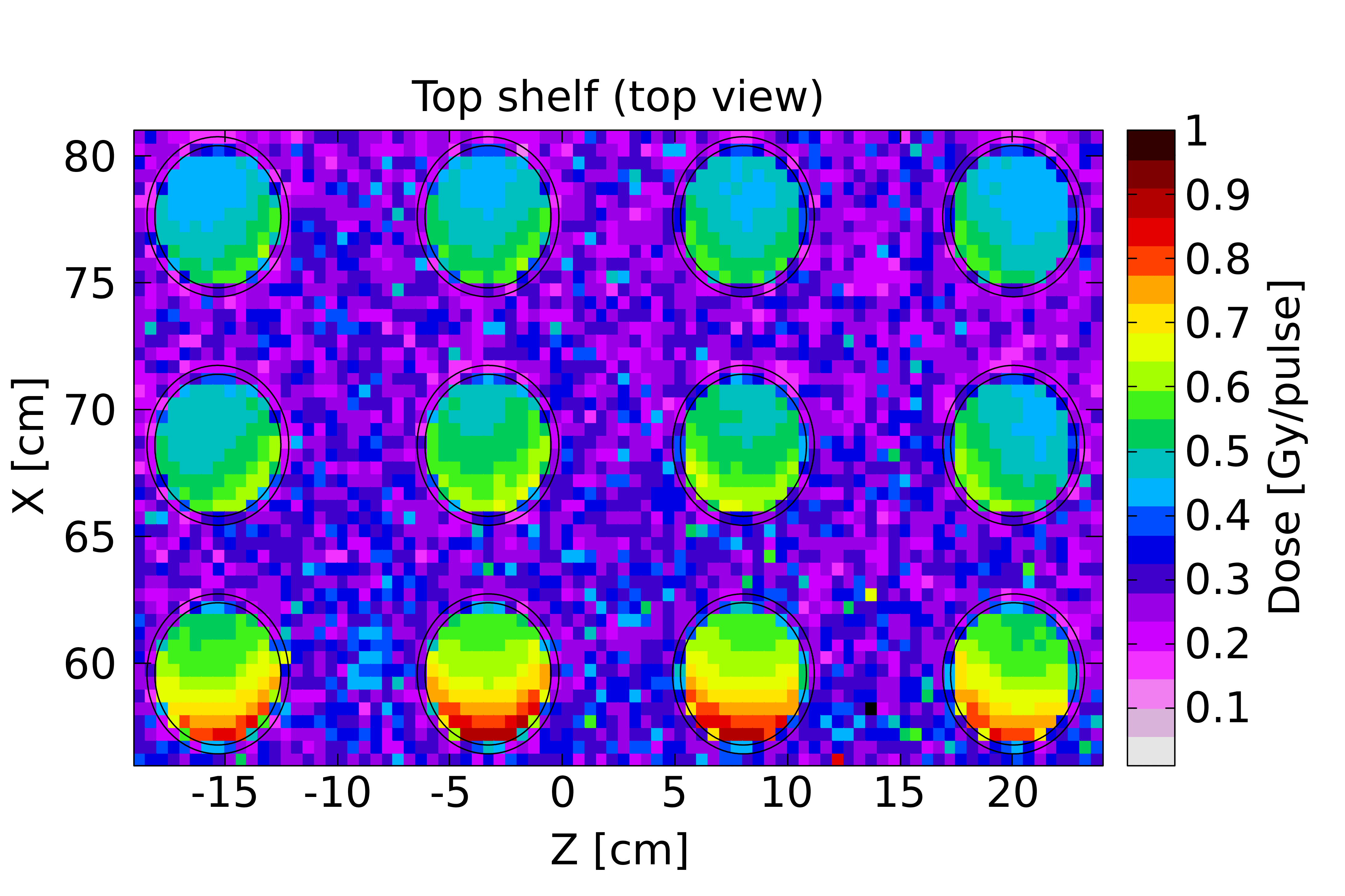}}
\subfigure[][]{\includegraphics[width=.45\textwidth]{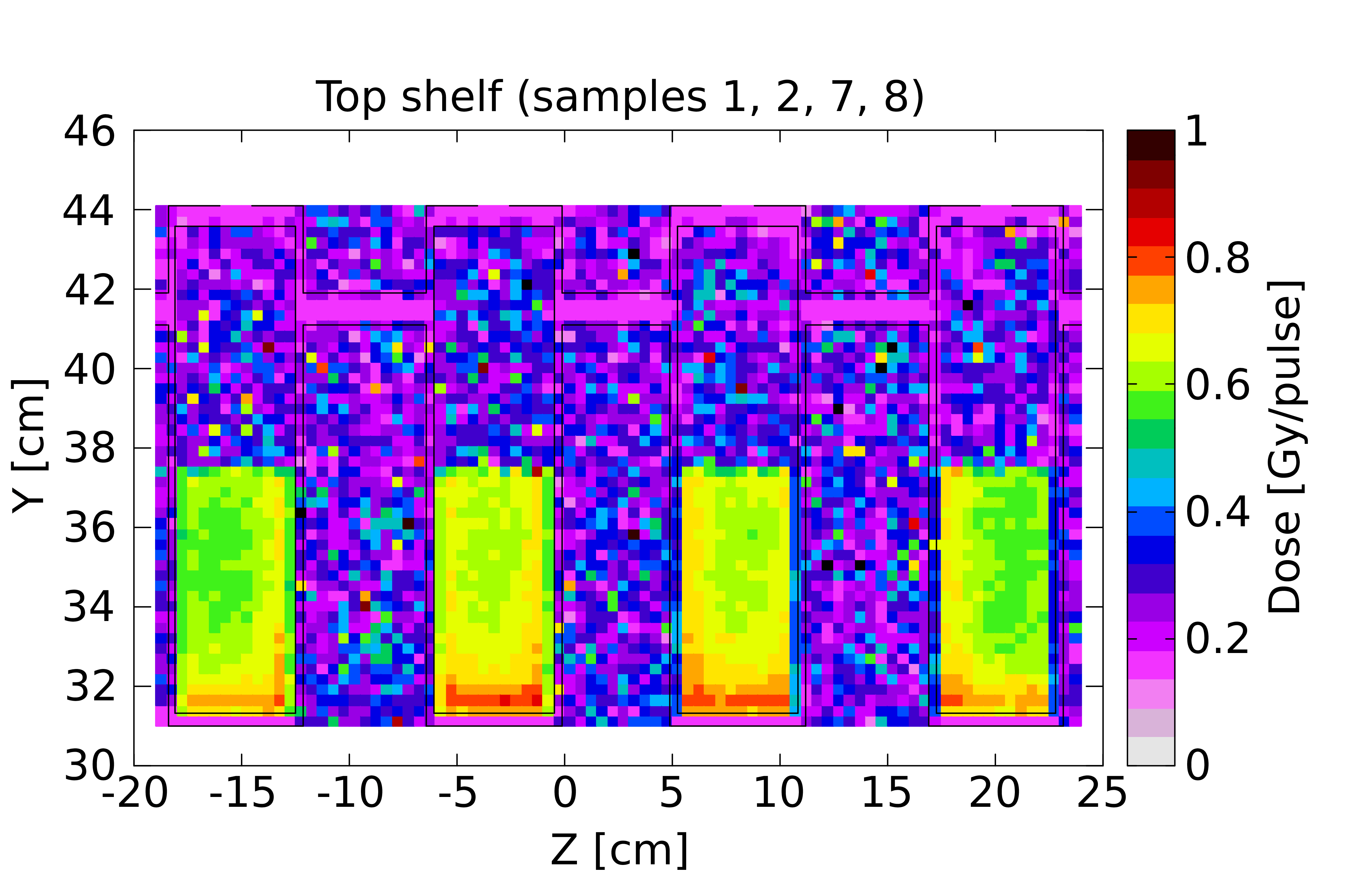}}
\end{center}
\caption{\label{fig:dose_mesh} Total dose in a selection of samples of the top shelf. a) Horizontal view of samples 1 to 12. b) Vertical view of samples 1, 2, 7, 8 (from left to right), having the same distance from the target. Dose values are reported in Gy/pulse.}
\end{figure}

The dose gradients within the sample volumes are clearly visible in Figure~\ref{fig:dose_mesh}, showing a mesh of total dose in the top shelf in different projections. The dose progressively decreases with the distance from the target along the x axis, while has higher homogeneity in the horizontal direction (z axis) and in the vertical one (y axis), as the distance from the target is roughly unchanged. Considering these gradients, an overall error of approx. $\pm$15\% on the total dose should be accounted for while averaging the dose on the whole sample volume.

\begin{table*}[htbp]
\centering
\caption{\label{tab:doses} Total dose in the samples, averaged over the four samples located at the same distance from the target, expressed in Gy/pulse. The neutron and gamma components of the dose are reported, as well as the ratio between the gamma component and the total. An overall uncertainty of about 15\% is attributed to the dose values.}
\smallskip
\begin{tabular}{lllllll}
\hline 
Shelf & Sample positions  & Total dose & n dose & $\gamma$ dose & $\gamma$ comp. \\
 & (average of four) & Gy/pulse & Gy/pulse & Gy/pulse & \% \\
\hline \hline

Top & 1, 2, 7, 8 & 0.68 & 0.55 & 0.13 & 19~\% \\
Top & 3, 4, 9, 10 & 0.54 & 0.41 & 0.13 & 24~\% \\
Top & 5, 6, 11, 12 & 0.50 & 0.38 & 0.12 & 24~\% \\
\hline
Bottom & 13, 14, 19, 20 & 0.45 & 0.32  & 0.13 & 29~\% \\
Bottom & 15, 16, 21, 22 & 0.38 & 0.25 & 0.13 & 34~\% \\
Bottom & 17, 18, 23, 24 & 0.36 & 0.24 & 0.12 & 33~\% \\

\hline \hline
\end{tabular}
\end{table*}

The discussed calculations highlight that neutron-dominated doses in the MGy range can be delivered in all the irradiation positions provided by the shelves, with a satisfactory homogeneity for macroscopic samples of up to 100~cm$^3$ of volume. 

Since the neutron dose, and accordingly the ratio between neutron and gamma component, are highly dependent on the total amount of hydrogen, variations in respect to the reported values are expected when different materials are irradiated. Dedicated calculations using refined compositions of the irradiated materials will be performed for future specific irradiation campaigns. 

\subsection{Particle energy spectra}
\label{sec:spectra}

\begin{figure*}[htbp!]
\centering 
\includegraphics[width=.8\textwidth]{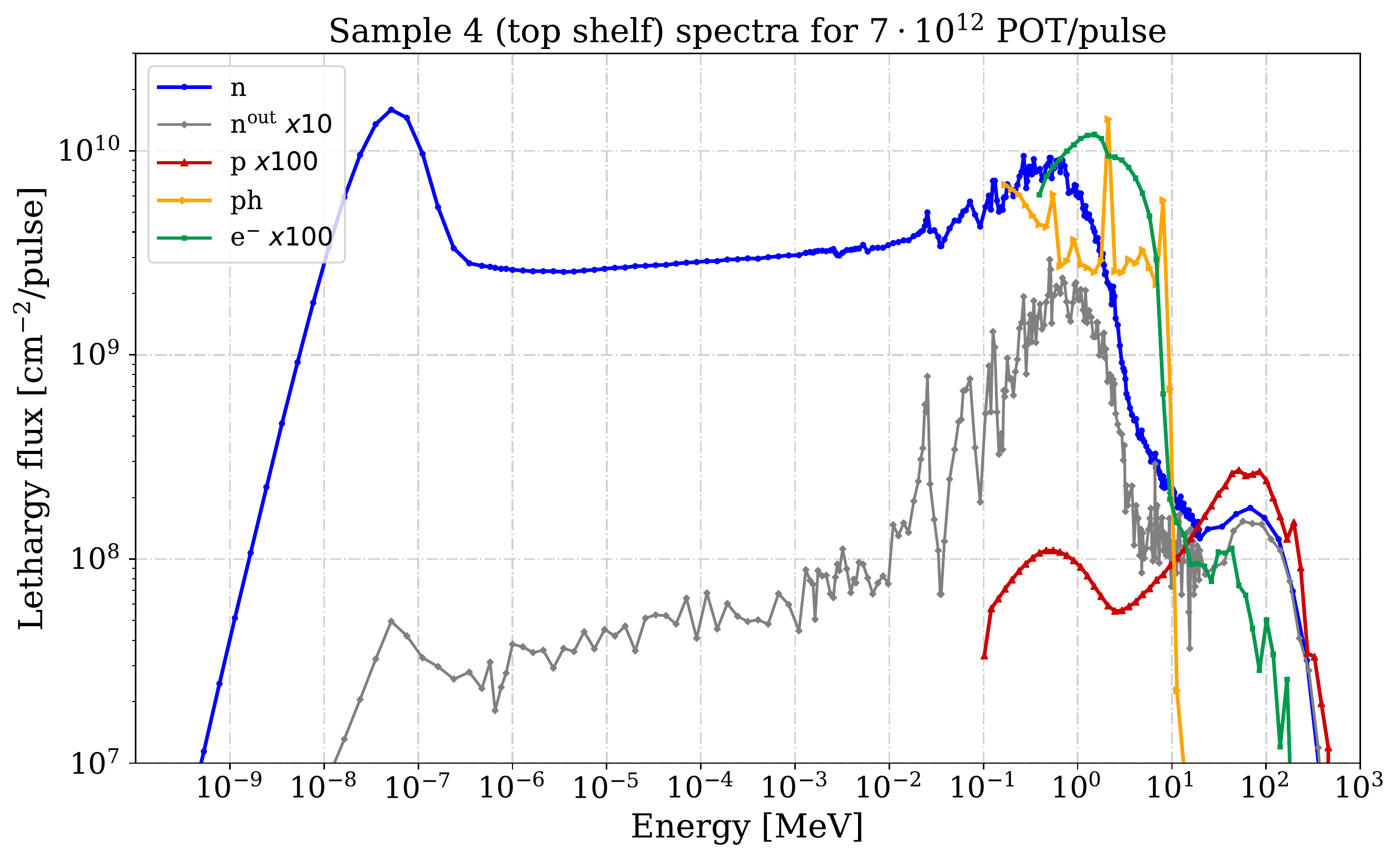}
\caption{\label{fig:spectrum} Particle spectra in a reference position of the top shelf. For comparison with the dominating neutrons, the proton and electron spectra are multiplied by a factor 100. As a matter of comparison, the grey curve represents the neutron fluence in a-NEAR (multiplied by 10).}
\end{figure*}

Figure~\ref{fig:spectrum} shows a typical neutron spectrum on one of the filled samples (number 4 of the top shelf) together with the proton, photon and electron spectra, whose fluences are several orders of magnitude lower compared to the neutron one. As can be seen, the neutron spectrum ranges from thermal up to about several GeV of maximum energy, and three main areas can be identified: thermal neutrons ($\sim$25~meV), evaporation ($\sim$1~MeV) and spallation ($\sim$100 MeV).

Other particles, whose contribution is orders of magnitude lower than the considered ones, such as muons and pions, have been neglected.

\subsection{Neutron fluence}
\label{sec:fluence}

Table~\ref{tab:fluences} reports neutron fluence values calculated by FLUKA in correspondence of irradiation position number 4 on the top shelf, as a general reference. The total neutron fluence is reported, as well as its thermal and fast components. The fast neutron fluence accounts for neutrons with energy higher than 0.5~MeV. High energy neutrons are expected to be the most effective in delivering dose to organic hydrogenated materials, as the main energy transfer process is represented by elastic scattering of neutrons on hydrogen nuclei. 

Thermal neutron fluence (n$_{\text{th}}$ in Table~\ref{tab:fluences}) is calculated following the standard used in electronics, accounting for all the neutrons with energies lower than 0.4~eV~\cite{JEDEC}. For completeness, thermal neutron equivalent fluence (n$_{\text{thEQ}}$) is reported in Table~\ref{tab:fluences} as well. It is particularly meaningful in the radiation to electronics context, especially when the sensitive volume of a device contains $^{10}B$. This quantity is calculated by convoluting the differential neutron flux with a weighting function, whose value is one at 25 meV and decreases as the inverse of the square root of the energy~\cite{cecchetto2018,garcia2017}.

In comparison to the fast neutrons, the contribution of thermal neutrons to the total neutron dose deposited in organic hydrogenated materials is expected to be negligible. However, it could become more relevant in the case of different compositions of the irradiated materials, for example when the composition includes in relevant amounts atoms with a high capture cross-section for thermal neutrons, such as $^{10}B$ isotope.

\begin{table*}[htbp]
\centering
\caption{\label{tab:fluences} Neutron fluence per pulse calculated by FLUKA in irradiation position 4 (top shelf). The total neutron fluence is reported, as well as the thermal neutron equivalent fluence (ThnEq), the thermal fluence based on JEDEC standard\cite{JEDEC} (ThN) and the fast neutron fluence. }
\smallskip
\begin{tabular}{lllllll}
\hline
n$_{\text{th}}$ & n$_{\text{thEQ}}$ (JEDEC $<$0.4eV) & n$_{\text{fast}}$ ($>$0.5 MeV) & Total \\
 n/cm$^2$/\text{pulse} & n/cm$^2$/\text{pulse}  & n/cm$^2$/\text{pulse} & n/cm$^2$/\text{pulse}  \\
\hline \hline 

2.8 $\cdot 10^{10}$ & 3.4 $\cdot 10^{10}$ & 1.0 $\cdot 10^{10}$& 9.3 $\cdot 10^{10}$ \\

\hline \hline 
\end{tabular}
\end{table*}

\section{Radiation protection considerations}
\label{sec:RP}

\subsection{Radiological situation of the n\_TOF NEAR and optimisation measures}
\label{sec:RP_NEAR}

Access to n\_TOF NEAR is only possible when the facility is not operational and is subject to RP authorisation, following a dedicated radiological risk assessment. Residual dose rates at a-NEAR when the shielding is closed are typically in the order of 100-300~$\mu$Sv/h after one day of cool-down time and of 10-30~$\mu$Sv/h after one week. However, access to the irradiation station requires the shielding to be open. The residual dose rate at i-NEAR can be in the order of several tens of mSv/h up to more than 100~mSv/h in close proximity to the spallation target. For this reason and according to the ALARA (As Low As Reasonably Achievable) principle, remote handling is deemed necessary to install and retrieve the samples, as described in Section~\ref{sec:handling}. This contributes to significantly reducing the dose received by personnel, as the robotic equipment can be controlled from a distance in low-dose-rate areas (a few $\mu$Sv/h).

\subsection{Estimates of samples' induced radioactivity}
In addition to the residual dose rate of the target area, the residual dose rate due to the induced activation of the samples is assessed; the results contribute to the definition of optimisation measures (such as additional cool-down time, shielding) in order to, first, safely retrieve and, then, handle the samples for post-irradiation analysis. Residual activation of samples and containers is estimated via FLUKA simulations. The following section provides an example of the RP studies performed for the first irradiation campaign started in 2021 (described in Section~\ref{sec:firstirradiations}).

\subsubsection{Simulation parameters}
The FLUKA simulations for RP purposes were performed implementing the geometry of the n\_TOF facility (as shown in Figure~\ref{fig:flukageo}) and selecting the value of 7.6~10$^{18}$~POT; this corresponds to the number of protons received by the n\_TOF target during the first irradiation campaign organised in 2021, which lasted about four months (described in Section~\ref{sec:firstirradiations}). However, the RP considerations reported here follow a conservative approach and  can also be considered as generally representative of future irradiation campaigns.

The residual dose rate was simulated for a set of six samples installed on their intermediate support in the top shelf, where the highest activation is expected. The chemical composition described in Table~\ref{tab:chemcomp} was used. As described in Section~\ref{sec:dosimetry}, this composition, including an estimation of traces of metallic impurities, is generally representative of the organic materials to be irradiated in the shelves. Further considerations on the specific materials to be irradiated are reported in Section~\ref{sec:firstirradiations}.

Different products or even the same product from a different batch generally have slightly different compositions and concentrations of metallic traces and therefore it is impossible to precisely predict their amount and the resulting activation.

\subsubsection{Residual dose rate}

\begin{figure}[htbp]
\centering 
\includegraphics[width=.45\textwidth]{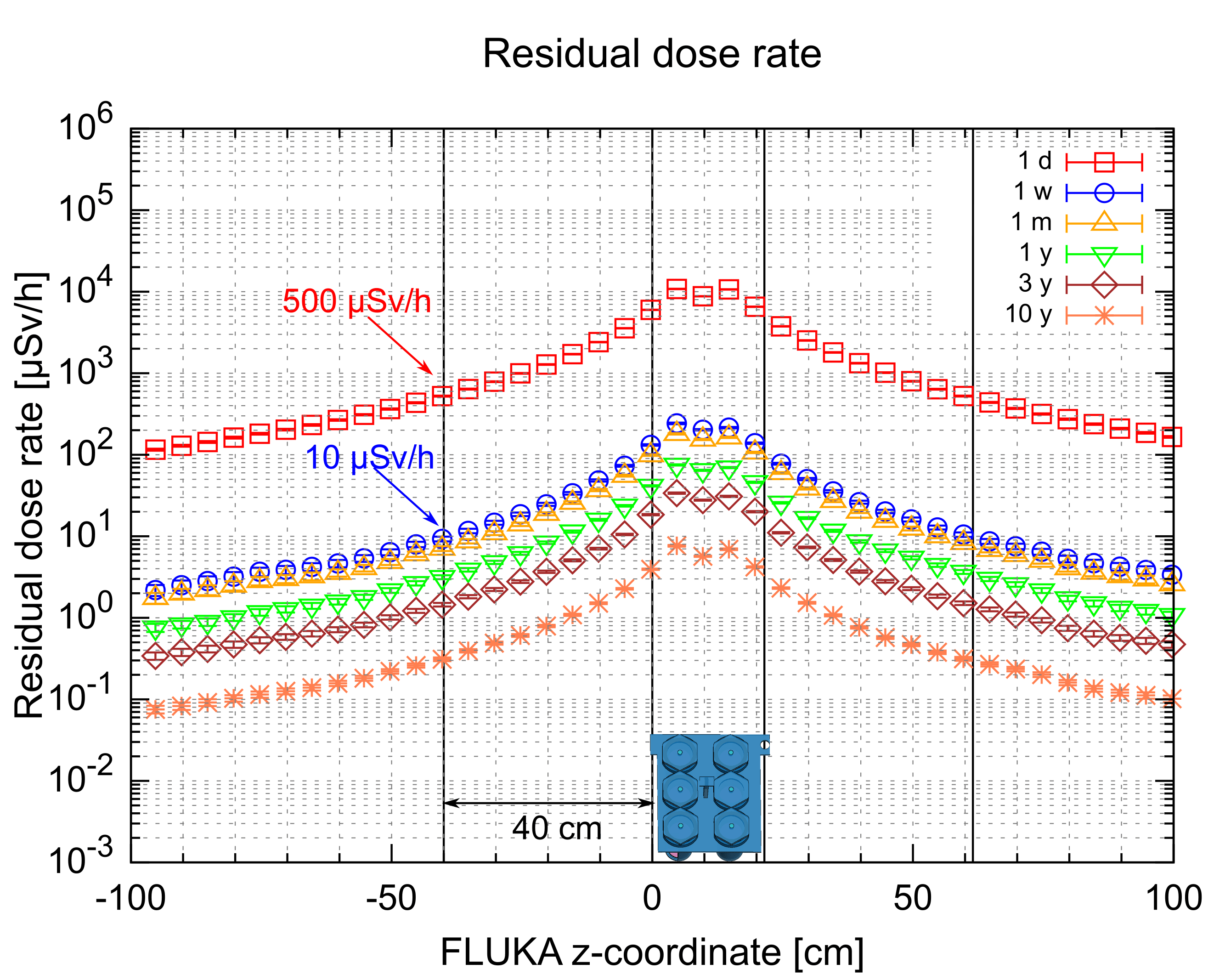}
\qquad
\caption{\label{fig:resdose1d} 1-D distribution along the FLUKA z-axis of the residual dose rate for various cool-down times (from 1 day up to 10 years).}
\end{figure}

Figure~\ref{fig:resdose1d} shows the residual dose rate as a function of the cool-down time (from 1 day up to 10 years) obtained from the FLUKA Monte Carlo simulations. The residual dose rate at 40~cm from the intermediate support containing a set of six containers and their samples reduces from about 500~$\mu$Sv/h after one day down to 10~$\mu$Sv/h after one week. This is related to the decay of Na-24 (half-life of about 15~hours), which is produced in the aluminium intermediate support, in the irradiated containers and in the samples (from Na traces). Based on these results and according to the ALARA principle, a minimum cool-down time of one week is implemented prior to the removal of the samples. A comparison between the simulated values and the results measured after the first irradiation is given in Section~\ref{sec:firstirradiations}. 

\subsubsection{Radionuclide inventory and radioactive waste aspects}

To evaluate the radionuclide inventory of a sample, the particle spectral fluences scored at the location of the samples were folded with nuclide production cross sections of the lubricant material by means of the ActiWiz analytical code~\cite{vincke2018}. By providing ActiWiz with the specific irradiation profile, the radionuclide inventory is obtained for the cool-down time of interest. The sample radio-toxicity is analysed in terms of the Swiss clearance limit (\emph{Limite de Libération} in French or LL), and the licensing limit, (\emph{Limite d’Autorisation} in French or LA. The inhalation of activity at the licensing limit on a single occasion yields a committed effective dose of 5~mSv.)~\cite{ORaP}. The LA indicates the value corresponding to the absolute activity level of a material above which handling of this material is subject to mandatory licensing and shall be performed in laboratories (called "Work Sectors") complying with specific requirements in terms of, e.g., radiation and fire protection, and ventilation.

Table~\ref{tab:LL} provides the fraction of LL and LA as a function of the cool-down time together with the main contributors, which are Na-22, Na-24 (only for a cool-down time of 1 day) and Zn-65 (originating from impurities of Na and Zn present in the samples). Even after ten years of cool-down time the fraction of LL for a sample is higher than one, which means that it cannot be cleared from regulatory control and it should therefore be disposed of as a radioactive waste. However, it should be noted that the final conclusion strongly depends on the amount of impurities; it cannot be excluded that very pure samples could become conventional waste after irradiation. This will be evaluated after the sample removal, combining the results from the FLUKA simulations with dedicated ~$\gamma$-spectrometry measurements.

\begin{table*}[htbp]
\centering
\caption{\label{tab:LL} Multiple of the clearance limit (limite de libération (LL), top) and of licensing limit (limite d’autorisation (LA), bottom) for the main radionuclides produced in the lubricant sample as a function of the cool-down time $t_{c}$ The relative contribution to the total LL is given as well.}
\smallskip
\begin{tabular}{llllllllll}
\hline
Nuclide & $t_{1/2}$  & $t_{c}$ & $t_{c}$& $t_{c}$ &  $t_{c}$& $t_{c}$ & $t_{c}$ \\
 &  & 1~d &1~week & 1~month & 1~y & 3~y & 10~y \\

\hline
 LL  & & &  &  &  &  \\

Na-22 & 2.6 y & 1.07 $\cdot 10^{3}$ & 1.07 $\cdot 10^{3}$ & 1.05 $\cdot 10^{3}$ & 8.42 $\cdot 10^{2}$ & 4.84 $\cdot 10^{2}$ & 7.52 $\cdot 10$ \\
      &    & $<$1\%  & 2\% & 2\% & 4\% & 17\% & 92\% \\
      
Na-24 & 15 h & 3.89 $\cdot 10^{5}$ & 4.82 $\cdot 10^{2}$ & $<$0.1 & $<$0.1 & $<$0.1 & $<$0.1 \\
      &    & 87\%  & $<$1\% & $<$1\% & $<$1\% & $<$1\% & $<$1\% \\
      
Zn-65 & 244 d & 5.32 $\cdot 10^{4}$ & 5.23 $\cdot 10^{4}$ & 4.90 $\cdot 10^{4}$ & 1.89 $\cdot 10^{4}$ & 2.83 $\cdot 10^{3}$ & 1.68 \\
     &    & 12\%  & 95\% & 96\% & 95\% &  82\% & 2\% \\
     
Total &   &  4.45 $\cdot 10^{5}$ & 5.50 $\cdot 10^{4}$ & 5.09 $\cdot 10^{4}$ & 1.98 $\cdot 10^{4}$ & 2.89 $\cdot 10^{3}$ & 8.21 $\cdot 10$ \\

\hline

LA  & & &  &  &  &  \\

Na-24 & 15 h & $<$0.1 & $<$0.1 & $<$0.1 & $<$0.1 & $<$0.1  & $<$0.1\\
      &    & 94\%  & 2\% & $<$1\% & $<$1\% & $<$1\% & $<$1\% \\
      
Zn-65 & 244 d & 0.3 & 0.3 & 0.2 & 0.1 & $<$1\% & $<$1\% \\
     &    & 6\%  & 90\% & 93\% & 95\% &  82\% & $<$1\% \\
     
Total &   &  4.6 & 0.3 & 0.3 & 0.1 & $<$1\% & $<$1\% \\

\hline
\end{tabular}
\end{table*}

After 1 week of cool-down the total activity in the sample is already less than 1~LA and therefore sample handling inside a laboratory classified as work sector according to ref.~\cite{ORaP} would not be required.

\section{First irradiation of materials}
\label{sec:firstirradiations}

\subsection{Pilot irradiation completed in 2021}

The irradiation station was successfully installed at n\_TOF in June~2021, and the first samples were installed for irradiation in time for the first operation of the new spallation target, in July~2021. Figure~\ref{fig:sample_handling} shows Telemax installing the first set of samples, in Figure~\ref{fig:irradstation} the four sets of samples installed in the NEAR Irradiation station before irradiation are shown. The irradiated materials represent a selection of commercial lubricants and elastomeric materials of interest for various applications at CERN. Details on the selected products and on the post-irradiation characterization of the samples will be described in dedicated publications. 

The first pilot irradiation allowed the technical and safety aspects of the developed methodology to be verified. Measurements of the residual activation of the containers and of the different irradiated materials were performed after irradiation. After about four months of irradiation (less than the nominal duration of 200 days due to 2021 being a target commissioning year, hence of shorter duration), according to the simulations discussed in Section~\ref{sec:dosimetry}, samples absorbed doses approximately ranging between 0.4~MGy and 0.8~MGy. These doses refer to a generic material composition which, based on previous experimental analyses, is considered to be roughly representative of organic materials. The dose values here reported are intended to be general estimations only. Dedicated simulations taking into account the specific composition and irradiation position of each sample will be performed and the results will be detailed in future works, along with the post-irradiation analyses. 

The first sets of samples were successfully removed from the irradiation facility using Telemax, confirming the safe design of the irradiation facility and the feasibility of the whole procedure. 

\subsection{Residual activity after irradiation}
The measured residual dose rates of the intermediate support and of its respective six containers and samples after the first irradiation confirm that the containers can be safely and easily handled by human operators after some weeks of cooling time. 

The residual dose rate of the most-activated intermediate support, measured after removal of the sample containers and after 23~days of cooling time was of 320~$\mu$Sv/h at contact, of 35~$\mu$Sv/h at 10~cm of distance and of 8~$\mu$Sv/h at 40~cm of distance. The residual dose rate of the most radioactive containers, including inner container and corresponding sample, were measured. Depending on the specific distance from target, the overall residual dose rate for single containers irradiated in the top shelf after 23~days of cooling time ranged between 70~$\mu$Sv/h and 110~$\mu$Sv/h at contact, between 7~$\mu$Sv/h and 10~$\mu$Sv/h at a  distance of 10~cm and of a few~$\mu$Sv/h at a distance of 40~cm. The order of magnitude of these measured values matches the simulated values discussed in Section~\ref{sec:RP}. 

These values fully confirm the possibility of safely following the whole life cycle of the irradiated samples and containers in agreement with RP principles.  

\subsection{Container opening}

After irradiation, samples were examined under a fume hood. Custom-designed reachers (as shown in Figure~\ref{fig:pilot}) are used to extract the inner container from the outer one, allowing a distance of 40~cm to be maintained between the operator and the container, if needed. The reachers also enable extraction of the inner container without spilling non solid samples. The whole handling system is designed to be managed remotely. However, after the first irradiation the dose rate was sufficiently low to allow both the inner and the outer containers to be directly handled by human operators. 

Grease and of oil samples were examined after irradiation. The grease samples maintained their position in the inner containers, as confirmed by the absence of spills and by the absence of contamination in the inner containers lids. A small fraction of the oil sample (some~ml) spilled out of the inner container, however it was successfully contained by the outer container, thus avoiding any contamination of the environment.

\subsection{Dose measurements: readout of RPL dosimeters}
\label{sec:dosimeters}

\begin{figure}[htbp]
\centering 
\includegraphics[width=.5\textwidth]{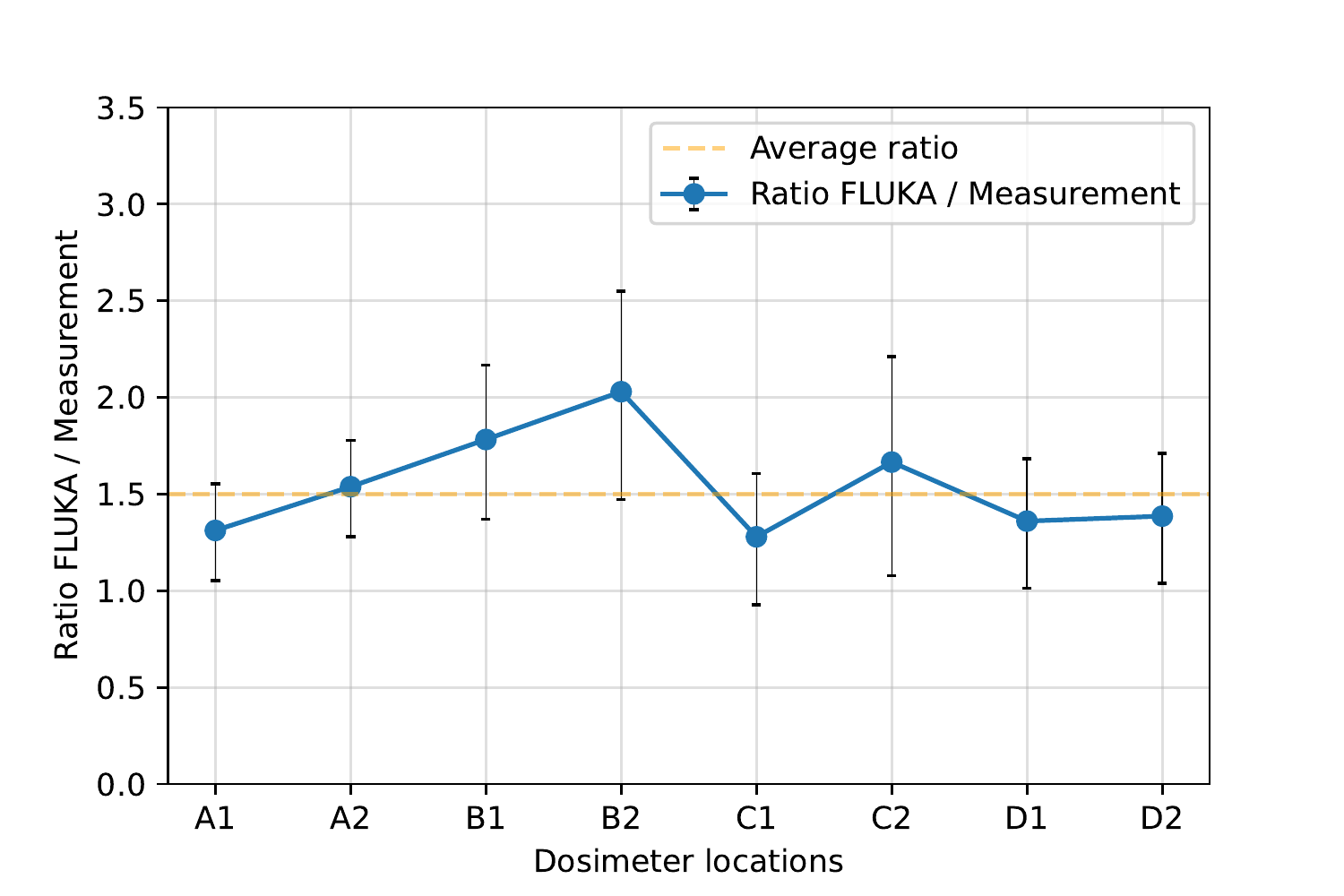}
\qquad
\caption{\label{fig:dosimeters} The figure shows the ratio between the simulated dose absorbed by the dosimeters and the experimental dose measurement, showing an overall good agreement within the experimental uncertainties.}
\end{figure}

As mentioned in Section~\ref{sec:design}, eight RPL dosimeters have been irradiated during the pilot irradiation completed in 2021. Figure~\ref{fig:dosimeter_position} shows the position of the dosimeters on the irradiation station, while Figure~\ref{fig:dosimeters} reports their readout measurement in comparison to the total dose simulated by FLUKA in the dosimeters themselves.

The doses measured by the dosimeters range from 100~kGy to 150~kGy, showing a satisfactory agreement with the simulated absorbed dose in the dosimeters, correspondingly ranging from 135~kGy to 200~kGy. The simulated doses systematically overestimate the measured ones by about 50\% on average, however, the measured and simulated values are overall compatible within the errors.

The dosimeters' readouts are roughly independent of the specific irradiation position within the irradiation station. By contrast, large differences are reported in the simulations of the dose absorbed by samples of organic materials in different positions, as discussed in Section~\ref{sec:dosimetry}. This can be explained considering the different composition of the dosimeters and of the samples, and the correspondingly different ratio between gamma and neutron dose. 

In the dosimeters, the neutron component of the dose represents on average 10\% of the total absorbed dose only, while in the organic samples it represents about 70-80\%. These differences are motivated by the much higher content of hydrogen and other light elements in organic samples, which highly influence the dose delivered by neutrons, due to moderation processes. This component is greatly reduced in the dosimeters, since they are made of phosphate glass and their composition is hydrogen free and carbon free. The gamma component of the dose, by contrast, is less dependent on these differences in the composition. In this scenario, the homogeneity of the dosimeter readouts can be explained. The dose gradients observed in the simulated samples, as reported in Table~\ref{tab:doses}, are in fact mostly due to the differences in the neutron component of the dose. 

This considered, the dose measured by the dosimeters can roughly be compared to the gamma component of the dose absorbed by the samples during the pilot irradiation, corresponding to about 140~kGy in all the positions, achieving a very satisfactory agreement.

The dose measured by the dosimeters can accordingly be considered as a good estimate of the gamma dose absorbed by organic materials in the irradiation station.

\subsection{Current and future irradiation studies on a selection of commercial materials}
The samples chosen for the first irradiation represent a selection of commercial products currently being studied and used at CERN, including both radiation tolerant and generic ones.

Currently, studies performed in the framework of the Radiation to Materials (R2M) activities focus on lubricants (in the form of greases and of fluid oils) and on elastomeric materials presently available in the market. In fact, they represent critical components necessarily used in high radiation areas~\cite{maestre2021}, and whose failure can have a high impact on accelerator operation~\cite{ferrari2021}. In particular, polyphenyl ether (PPE)-based lubricants are being irradiated and studied in detail, as their composition is known to be promising for radiation tolerance~\cite{bolt,patent, yellow3}. Recent studies have confirmed the superior resistance of PPE-based products~\cite{ferrari2019} and have evidenced possible differences between gamma and reactor mixed field in determining radiation damage \cite{ferrari2021oil}. The data collected at n\_TOF NEAR in the first pilot irradiation and in future irradiation campaigns will contribute in further understanding radiation damage mechanisms.

\section{Conclusions}
\label{sec:conclusions}

Construction and installation of a new CERN multi-purpose irradiation station was successfully completed during July 2021 in the  n\_TOF NEAR area. The new irradiation station takes advantage of the spallation neutrons produced by the recently installed third-generation n\_TOF spallation target and has minimal impact on the facility operation and physics programme. 

As a result, CERN is now equipped with several irradiation positions for materials, where unprecedented radiation damage data can be collected in a neutron-dominated environment. The available irradiation conditions are - for the first time - much closer, in comparison with standard gamma irradiations, to the ones encountered during operation in several accelerator facilities worldwide.

The data produced thanks to this new station will allow a better assessment of the radiation tolerance of commercial and custom-made materials, so far largely lacking or undocumented, and permit a corresponding increase of the lifetime of devices used in high-radiation areas, thereby reducing the risk of failures during operation of accelerators and other facilities with similar challenges.

Radiation safety is a key aspect in the design and operation of the irradiation station and post-irradiation analysis of samples. Simulations, including FLUKA Monte Carlo techniques, were extensively used to predict dosimetry in the samples and the resulting induced activation. Remote handling via telemanipulated robots was integrated into the design and operating procedures at an early stage and extensive mock-up testing was used to demonstrate the safety of the whole sample life cycle. Sample containers and their supports were subjected to in-depth risk analysis and practical testing before obtaining approval to carry out the first pilot irradiation. 

Successful pilot irradiation of material samples over a four month period and subsequent inspections and measurements on the irradiated samples demonstrated the effectiveness of the design of the irradiation station and its sample containers along with the safe and reliable methodology for sample irradiation and handling.

\section*{Acknowledgments}
The authors would like to acknowledge the support of CERN's Sources, Targets and Interactions (STI) Group, the Radiation-to-Electronics (R2E) Project as well as the Accelerator Consolidation (ACC-CONS) Project at CERN. The authors would like to express thanks to S.~Sorlut for the design of the NEAR collimator as well as to D.~Boon for the construction of the target mobile shielding.

\bibliography{bibfile}
\bibliographystyle{unsrturl}

\end{document}